\documentclass[11pt,a4paper]{article}
\usepackage{graphicx}
\usepackage{amsmath}
\usepackage{amssymb}
\usepackage{bm}
\usepackage{doi}
\usepackage[top=12mm,bottom=12mm,left=30mm,right=30mm,head=12mm,includeheadfoot]{geometry}
\numberwithin{equation}{section}
\usepackage{hyperref}

\begin{document}

\begin{center}{\Large \textbf{
\emph{Tangent fermions:}
Dirac or Majorana fermions\smallskip\\
on a lattice without fermion doubling}}\end{center}
\begin{center}
\textbf{C.W.J. Beenakker,$^{\mathbf{1}}$ A. Don\'{i}s Vela,$^{\mathbf{1}}$ G. Lemut,$^{\mathbf{1}}$\smallskip\\
M.J. Pacholski,$^{\mathbf{1,2}}$ and J. Tworzyd{\l}o$^{\mathbf{3}}$}
\end{center}
\begin{center}
\textbf{1} Instituut-Lorentz, Universiteit Leiden, P.O. Box 9506,\\ 
2300 RA Leiden, The Netherlands\\
\textbf{2} Max Planck Institute for the Physics of Complex Systems,\\
N\"{o}thnitzer Strasse 38, 01187 Dresden, Germany\\
\textbf{3} Faculty of Physics, University of Warsaw, ul.\ Pasteura 5,\\
02--093 Warszawa, Poland
\end{center}
\begin{center}
(February 2023)
\end{center}
\section*{Abstract}
\textbf{
We review methods to discretize the Hamiltonian of a topological insulator or topological superconductor, without giving up on the topological protection of the massless excitations (respectively, Dirac fermions or Majorana fermions). The method of tangent fermions, pioneered by Richard Stacey, is singled out as being uniquely suited for this purpose. Tangent fermions propagate on a $\mathbf{2\bm{+}1}$ dimensional space-time lattice with a tangent dispersion: $\mathbf{tan^2 (\bm{\varepsilon}/2) \bm{=} tan^2 (k_x/2) \bm{+}tan^2 (k_y/2)}$ in dimensionless units. They avoid the fermion doubling lattice artefact that would spoil the topological protection, while preserving the fundamental symmetries of the Dirac Hamiltonian. Although the discretized Hamiltonian is nonlocal, as required by the fermion-doubling no-go theorem, it is possible to transform the wave equation into a generalized eigenproblem that is local in space and time. Applications that we discuss include Klein tunneling of Dirac fermions through a potential barrier, the absence of localization by disorder, the anomalous quantum Hall effect in a magnetic field, and the thermal metal of Majorana fermions.
}

\vspace{10pt}
\noindent\rule{\textwidth}{1pt}
\tableofcontents
\noindent\rule{\textwidth}{1pt}
\vspace{10pt}

\section{Introduction}
\label{intro}

Topological insulators have a gapped interior and gapless excitations on the surface, described by the Dirac Hamiltonian $H=\hbar v\bm{k}\cdot\bm{\sigma}$. These are massless quasiparticles, with a cone-shaped relativistic dispersion relation $E=\pm \hbar v|\bm{k}|$. For computational purposes one would like to discretize $H$, replacing the momentum operator $\bm{k}=-i\partial/\partial_{\bm r}$ by a finite difference on a two-dimensional (2D) lattice. One then runs into a lattice artefact known as fermion doubling: A spurious additional species of low-energy excitations appears, no matter how small the lattice constant is.

\begin{figure}[tb]
\centerline{\includegraphics[width=0.5\linewidth]{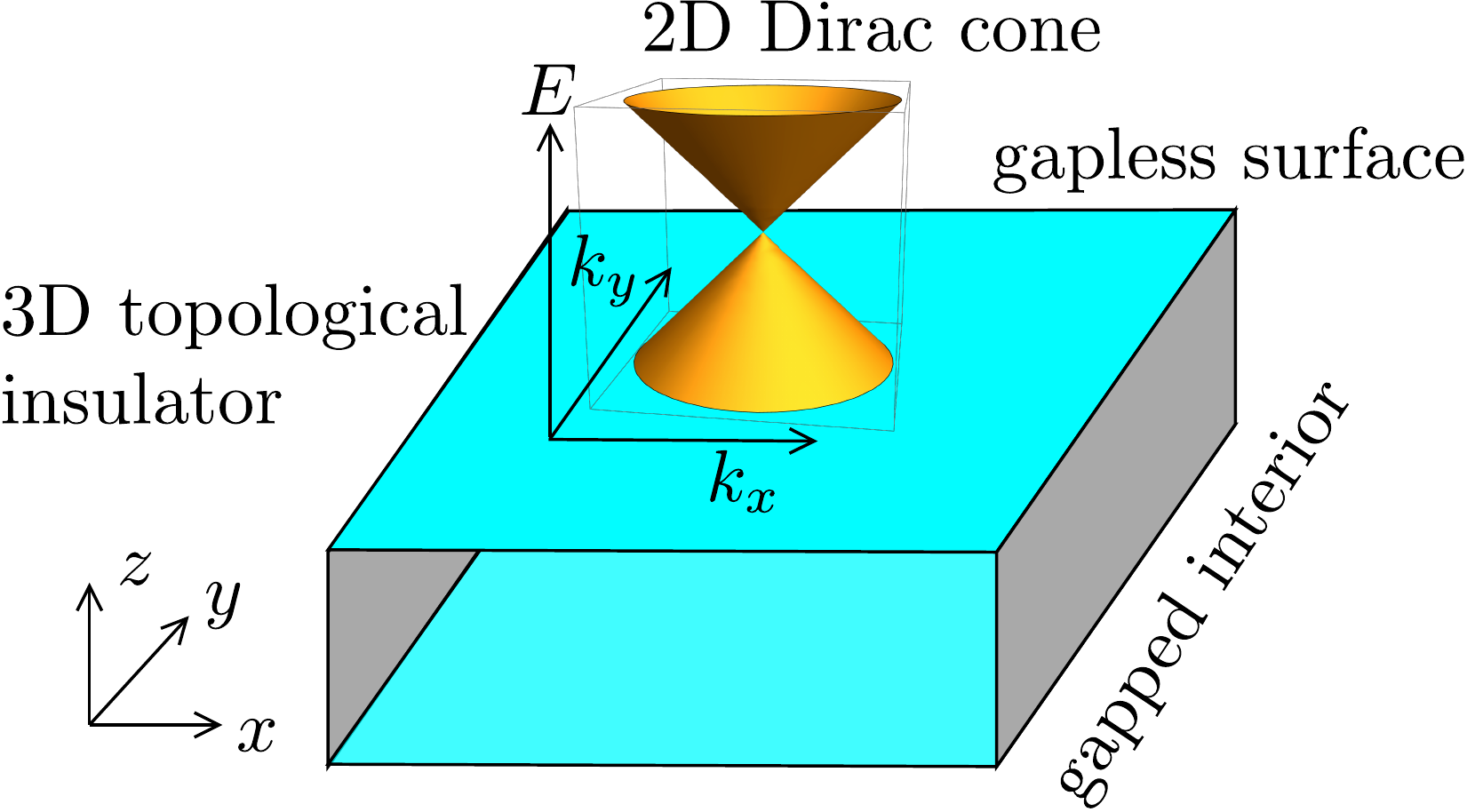}}
\caption{A topological insulator has a gapped interior and a gapless surface. Massless Dirac fermions exist on the surface, with a conical dispersion $E(k_x,k_y)$ (assuming a surface in the $x$--$y$ plane of infinite extent). If the surface Hamiltonian is discretized additional Dirac cones may appear at the edge of the Brillouin zone (fermion doubling). We review methods to avoid this lattice artefact.
}
\label{fig_layout}
\end{figure}

Fermion doubling is a notorious complication in particle physics \cite{Tong}, governed by the no-go theorem of Nielsen and Ninomiya \cite{Nie81}: Any local Hamiltonian that preserves the chirality of the Dirac fermions must have an even number of conical points in the Brillouin zone. One of the work-arounds invented in that context \cite{Kap92} is actually the one chosen by Nature in a topological insulator: The 2D lattice is embedded in a 3D lattice, say in a slab geometry (see Fig.\ \ref{fig_layout}). One can then think of the Dirac cone on the top surface as being doubled on the bottom surface, but if the surfaces are widely separated there will effectively be only a single species of massless excitations on each surface.

Since it is computationally costly to work with a 3D lattice \cite{Zie23}, a fully 2D formulation is preferrable. In what follows we will review the options developed by particle physicists, with one key criterion in mind: If we add disorder to the Dirac Hamiltonian, as is unvoidable in a real material, will the Dirac cone remain gapless? The robustness is known as ``topological protection'', it is the defining characteristic of a topological insulator \cite{Chi16}. This criterion has not played a decisive role in particle physics, presumably because disorder is not a relevant ingredient in that context.
 
We have found one method that satisfies the criterion of avoiding fermion doubling while preserving topological protection. It has its roots in the particle physics literature \cite{Sta82,Ben83}, but has not been widely adopted by that community.\footnote{Stacey's 1982 paper \cite{Sta82} received only about 20 citations (less than one per year) before it was noticed by condensed matter physicists in 2008 \cite{Two08}.} Because the linear dispersion relation is replaced by a tangent, we will refer to it as the method of \textit{tangent fermions}.

In Table \ref{table_symm} we summarize the properties of the various discretization schemes that we will discuss. Each discretization has its own dispersion relation, which reduces to the linear dispersion near the physical Dirac point at the center $\bm{k}=0$ of the Brillouin zone. The distinguishing properties include: 
\begin{itemize}
\item the symmetries that the discretization does or does not preserve --- the chiral symmetry which defines the handedness of the particles and the symplectic symmetry which is the time-reversal symmetry for spin-$1/2$ particles;
\item the number of Dirac points in the Brillouin zone (1 if there is no fermion doubling);
\item the locality of the discretization, meaning whether the discretized Hamiltonian only couples nearby lattice points; 
\item and finally the presence or absence of the protection against gap opening by disorder.
\end{itemize}

\begin{table}[tb]
\begin{center}
\begin{tabular}{c|c|c|c|c|c}
&chiral&symplectic&&&topological\\
dispersion&symm.&symm.&Dirac points&locality&protection\\
\hline
sine &$\checkmark $&$\checkmark $&4&$\checkmark $&$\times$\\
sine+cosine (Wilson \cite{Wil74})&$\times$&$\times$&1&$\checkmark$&$\times$\\
staggered (Susskind \cite{Sus77})&$\checkmark$&$\times$&2&$\checkmark$&$\times$\\
linear sawtooth ({\sc slac} \cite{Dre76})&$\checkmark$&$\checkmark$&1&$\times$&$\times$\\
tangent (Stacey \cite{Sta82})&$\checkmark$&$\checkmark$&1&$\times (\checkmark)$&$\checkmark$\\
\end{tabular}
\end{center}
\caption{Five approaches to discretize the Dirac equation on a 2D lattice. The presence or absence of a property is indicated by $\checkmark$ or $\times$, respectively. The tangent dispersion has a nonlocal Hamiltonian, but it allows a local formulation of a generalized eigenproblem (hence the $\checkmark$ in parentheses). Only the tangent dispersion has an unpaired and topologically protected Dirac point.
}
\label{table_symm}
\end{table}

The outline of this review is as follows. In the next two sections we first introduce the fermion-doubling problem of massless lattice fermions, and then compare the various approaches that have been proposed to overcome this problem. The approach of tangent fermions is particularly promising for applications in topological quantum matter, because it preserves the topological protection of the massless quasiparticles. We review the features of that approach from the perspective of a characteristic set of applications in topological insulators and topological superconductors. Computer codes for the various applications are made available in a repository \cite{zenodo}.

\section{Two-dimensional lattice fermions}
\label{sec_2Dlattice}

A Dirac point $\bm{q}$ in 2D momentum space is a crossing point of two energy bands, of the form 
\begin{equation}
E(\bm{k})=E_0\pm\hbar v |\bm{k}-\bm{q}| +{\cal O}(\bm{k}-\bm{q})^2.
\end{equation}
On a 2D lattice the momenta can be restricted to a compact region, the Brillouin zone, spanned by reciprocal lattice vectors. Momenta related by a reciprocal lattice vector are equivalent. No fermion doubling means that there is a single inequivalent Dirac point in the Brillouin zone, in other words, all Dirac points in momentum space are related by reciprocal lattice vectors.

In the absence of any symmetry the Dirac point is unstable: A small perturbation may open a gap at $\bm{k}=\bm{q}$, converting the crossing into an anti-crossing. Crystalline symmetries can stabilize a Dirac point \cite{Le22}. One speaks of topological protection if a gap opening is prevented irrespective of the crystal structure.

Topological protection relies on the presence of either chiral symmetry or symplectic symmetry \cite{Chi16}. \textit{Chiral symmetry} requires that the Hamiltonian anticommutes with a unitary operator. One can then associate a winding number $\pm 1$ with a Dirac point. A gap opening would imply a discontinuous change to zero of the winding number, which cannot happen in response to a small perturbation of the Hamiltonian. \textit{Symplectic symmetry} is the time-reversal symmetry of a spin-$1/2$ particle.\footnote{Symplectic symmetry is broken by magnetic field, but this is not the only way. In a thin-film geometry the coupling of top and bottom surfaces introduces a term in the effective surface Hamiltonian that breaks the symplectic symmetry --- in the absence of any magnetization.} It enforces a two-fold degeneracy of the energy levels (Kramers theorem), which prevents a crossing of two bands from evolving into an anticrossing.

Only \textit{unpaired} Dirac points are topologically protected: Dirac cones may gap out pairwise without changing the net winding number or without violating Kramers degeneracy. This is a major obstacle, because Dirac cones tend to appear in pairs on a lattice.

If both chiral and symplectic symmetry are maintained, the Brillouin zone contains a Dirac point at zero energy at each momentum $\bm{q}$ which differs from $-\bm{q}$ by a reciprocal lattice vector. There are $2^d$ such time-reversally invariant momenta in $d$ dimensions, so 4 in 2D. If we then break symplectic symmetry we can move the Dirac points around and gap them out pairwise by merging two Dirac cones with opposite winding number. However, we can not end up with an unpaired Dirac cone unless we also break chiral symmetry --- spoiling the topological protection.

This obstruction to unpaired Dirac cones in a 2D system can be expressed by a no-go theorem:\footnote{This is a stronger statement than the Nielsen-Ninomiya no-go theorem for a 3D system, which only requires chiral symmetry breaking.} \textit{A local discretization of the 2D Dirac Hamiltonian cannot have an unpaired Dirac cone, unless it breaks both chiral and symplectic symmetries.} The ``locality'' condition provides a work around: a nonlocal discretization can have discontinuities or poles in the dispersion relation, which may ``hide'' a Dirac point. One can check that the entries in Table \ref{table_symm} are consistent with this no-go theorem.

\section{Methods to discretize the Dirac equation}
\label{sec_methods}

We now turn to the overview of methods to discretize the 2D Dirac Hamiltonian,
\begin{equation}
H_{0}=\hbar v(k_x\sigma_x+k_y\sigma_y)=\hbar v\begin{pmatrix}
0&-i\partial_x-\partial_y\\
-i\partial_x+\partial_y&0
\end{pmatrix},\label{H0def}
\end{equation}
focusing first on the case that the massless electrons can move freely on the $x$--$y$ plane, without any electromagnetic fields. The Dirac fermions have energy independent velocity $v$. The Pauli spin matrices $\bm{\sigma}$ are coupled to the momentum $\bm{k}=-i\partial_{\bm{r}}$. In Eq.\ \eqref{H0def} the spin-momentum locking is such that the spin points parallel to the momentum. The alternative perpendicular spin-momentum locking ($k_x\sigma_y-k_y\sigma_x$) can be obtained by a unitary transformation of $H_0$, so we need not distinguish the two cases here.

The energy-momentum relation (dispersion relation) of the Dirac Hamiltonian,
\begin{equation}
E(\bm{k})^2 = (\hbar v)^2(k_x^2+k_y^2),\label{dispersion}
\end{equation}
consists of a pair of cones that touch at the point $\bm{k}=0$ --- the Dirac point. When the Hamiltonian is discretized on a lattice the dispersion relation becomes periodic: $E(\bm{k}+\bm{K})=E(\bm{k})$ for any reciprocal lattice vector $\bm{K}$. Momenta which are not related by a reciprocal lattice vector form the Brillouin zone. For some discretization methods the Dirac point at $\bm{k}=0$ is copied at other points in the Brillouin zone (fermion doubling).

The Dirac Hamiltonian \eqref{H0def} satisfies the two symmetry relations introduced in the previous section,
\begin{equation}
\label{symmetries}
\begin{split}
\text{chiral symmetry:}&\;\;\sigma_z H_0\sigma_z=-H_0,\\
\text{symplectic symmetry:}&\;\; \sigma_y H_0^\ast\sigma_y=H_0.
\end{split}
\end{equation}
The complex conjugation is taken in the real-space basis, so the sign of both momentum and spin is inverted by the symplectic symmetry operation. For each discretization method we will check whether the symmetries \eqref{symmetries} are preserved or not.

The topological protection of the Dirac point relies on the absence of fermion doubling and on the conservation of at least one of the two fundamental symmetries \eqref{symmetries}. The linearity of the dispersion relation, $E\propto|\bm{k}|$, may be a desirable feature, but it is not essential for the protection. What is essential for a practical method is that the eigenvalue problem can be solved using linear algebra of sparse matrices. This is the issue of locality of the discretization.

\subsection{Sine dispersion}

We start with a square lattice, lattice constant $a$, and discretize the derivative operator by the first order finite difference:
\begin{equation}
\partial_x f(x,y)\mapsto (2a)^{-1}[f(x+a,y)-f(x-a,y)],\label{sinediscrete}
\end{equation}
and similarly for $\partial_y f(x,y)$. Notice that $e^{a\partial_x}=e^{iak_x}$ is the translation operator, $e^{a\partial_x}f(x)=f(x+a)$. The discretization \eqref{sinediscrete} therefore gives the Hamiltonian
\begin{equation}
H_{\rm sine}=(\hbar v/a)(\sigma_x\sin ak_x+\sigma_y\sin ak_y),
\end{equation}
with the sine dispersion
\begin{equation}
E_{\rm sine}(\bm{k})^2=(\hbar v/a)^2(\sin^2 ak_x+\sin^2 ak_y).
\end{equation}

Chiral symmetry and symplectic symmetry \eqref{symmetries} are both preserved by  the Hamiltonian $H_{\rm sine}$, but there is fermion doubling: In the Brillouin zone $|k_x|<\pi/a$, $|k_y|<\pi/a$ there are Dirac points at each of the time-reversally invariant momenta: the center $\bm{k}=0$, the corners $|k_x|=|k_y|=\pi/a$ and the midpoints $k_x=0,|k_y|=\pi/a$, $k_y=0,|k_x|=\pi/a$. The four corners and opposite midpoints are related by a linear combination of reciprocal lattice vectors $\bm{K}=(2\pi/a,0)$ and $\bm{K}'=(0,2\pi/a)$, so there are 4 inequivalent Dirac points in the Brillouin zone.

\subsection{Sine plus cosine dispersion}

An effective way to remove the spurious Dirac points is to gap them by the addition of a momentum dependent magnetization $\mu(\bm{k})\sigma_z$ to the Dirac Hamiltonian. If $\mu$ vanishes at $\bm{k}=0$ the physical Dirac point at the center of the Brillouin zone is unaffected. This is the approach introduced by Wilson \cite{Wil74,Gin82}. A quadratic $\mu\propto k^2$ is discretized on a square lattice, resulting in the Hamiltonian
\begin{equation}
H_{\rm Wilson}=(\hbar v/a)(\sigma_x\sin ak_x+\sigma_y\sin ak_y)+m_0\sigma_z(2-\cos ak_x-\cos a k_y),\label{HWilson}
\end{equation}
with the sine plus cosine dispersion
\begin{equation}
E_{\rm Wilson}(\bm{k})^2=(\hbar v/a)^2(\sin^2 ak_x+\sin^2 ak_y)+m_0^2(2-\cos ak_x-\cos a k_y)^2.\label{EWilson}
\end{equation}
The Dirac points of the sine dispersion acquire a gap $\propto m_0$, only the Dirac point at $\bm{k}=0$ remains gapless. 

Fermion doubling in Wilson's approach is avoided at expense of a breaking of both chiral and symplectic symmetries. The product of these two symmetries is preserved,
\begin{equation}
\sigma_x H_{\rm Wilson}^\ast\sigma_x=-H_{\rm Wilson}, 
\end{equation}
which is sufficient for some applications \cite{Hon12,Mas15,Mes17,Ara19}.

\subsection{Staggered lattice dispersion}

\begin{figure}[tb]
\centerline{\includegraphics[width=0.5\linewidth]{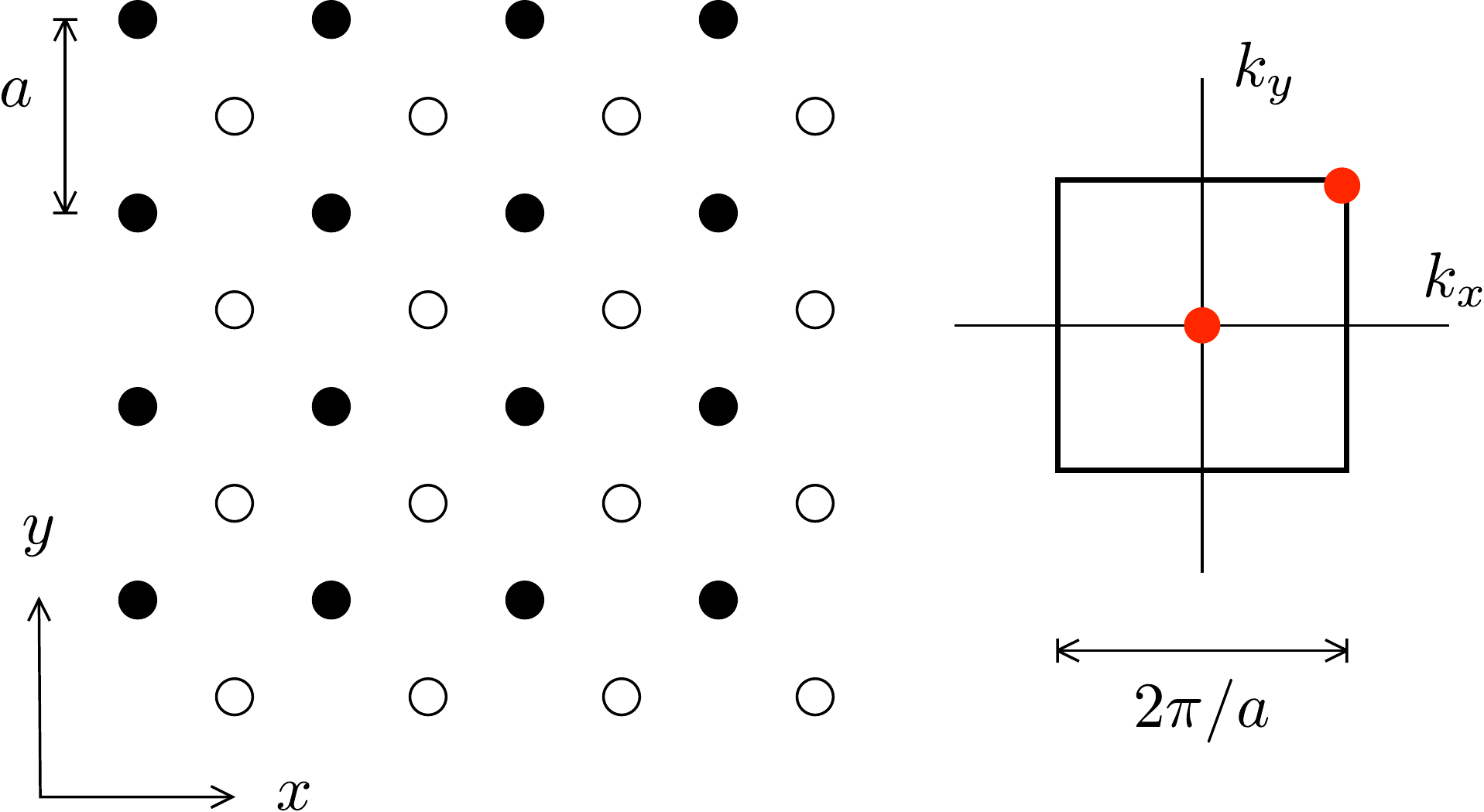}}
\caption{Left panel: Staggered pair of grids for the discretization of Dirac fermions in Susskind's approach. The black and white dots distinguish the $u$ and $v$ amplitudes of the spinor wave function $\Psi=(u,v)$. Right panel: The square shows the Brillouin zone in momentum space, the red dots indicate two inequivalent Dirac points.
}
\label{fig_grids}
\end{figure}

Much of the particle physics literature follows Susskind's approach \cite{Kog75,Sus77}, which applies a different lattice to each of the two components of the spinor wave function $\Psi=(u,v)$. The two lattices are staggered, see Fig.\ \ref{fig_grids}, displaced by half a lattice constant. The momentum operator transfers from one lattice to the other, which amounts to a diagonal displacement by a distance of $a/\sqrt 2$, as expressed by the translation operators $e^{ia(k_x\pm k_y)/2}$.

The discretized Dirac Hamiltonian still acts on the original lattice (black dots in Fig.\ \ref{fig_grids}). The unitary transformation with operator
\begin{equation}
U_{\rm stagger}=\begin{pmatrix}
1&0\\
0&e^{ia(k_x+k_y)/2}
\end{pmatrix}
\end{equation}
initializes the pair of staggered lattices ($u$ component on the black dots, $v$-component on the white dots). The Hamiltonian then takes the form 
\begin{equation}
H_{\rm Susskind}=\sqrt{2}(\hbar v/a)U_{\rm stagger}^\dagger\bigl(\sigma_x\sin[a(k_x-k_y)/2]+\sigma_y\sin[a(k_x+k_y)/2]\bigr)U_{\rm stagger}.\label{HSusskind}
\end{equation}
Check that the $2\pi/a$ periodicity in the $k_x$ and $k_y$ components is maintained: the minus sign picked up by the sine terms is canceled by the unitaries.

In terms of the rotated momenta $q_x=(k_x-k_y)/\sqrt 2$, $q_y=(k_x+k_y)/\sqrt 2$, normalized such that $|\bm{q}|^2=|\bm{k}|^2$, one has
\begin{equation}
H_{\rm Susskind}=\hbar v[q_x\sigma_x+q_y\sigma_y+{\cal O}(q^2)],
\end{equation}
so the Dirac Hamiltonian \eqref{H0def} is recovered in the continuum limit.

The corresponding dispersion relation
\begin{equation}
E_{\rm Susskind}(\bm{k})^2=2(\hbar v/a)^2\bigl(\sin^2[(k_x-k_y)/2]+\sin^2[(k_x+k_y)/2]\bigr)
\end{equation}
has two inequivalent Dirac points in the Brillouin zone, at the center and at the corner. Compared to the sine discretization the staggered lattice has reduced the number of Dirac points from four to two, but fermion doubling has not been fully eliminated. Chiral symmetry is preserved, but symplectic symmetry is broken by the relative displacement of the two spinor components.

More generally, on a $d$-dimensional lattice the sine dispersion has $2^d$ inequivalent Dirac points in the Brillouin zone (one at each time-reversally invariant momentum), and the staggered lattice reduces that by one half. For $d=1$ this is sufficient to avoid fermion doubling. In that case the Susskind Hamiltonian \eqref{HSusskind} is equivalent (up to a unitary transformation) to the 1D Wilson Hamiltonian
\begin{equation}
H_{\rm Wilson}(k_x,k_y=0)=(\hbar v/a) \sigma_x\sin ak_x+m_0\sigma_z(1-\cos ak_x)
\end{equation}
for the special value $m_0=\hbar v/a$. The resulting $\sin(ak_x/2)$ dispersion is shown in Fig.\ \ref{fig_dispersion} (green curve).

\subsection{Linear sawtooth dispersion}

The discretization schemes discussed in the previous subsection are all local, in the sense that they produce a sparse Hamiltonian: each lattice site is only coupled to a few neighbors. If one is willing to abandon the locality of the Hamiltonian, one can eliminate the fermion doubling by a discretization of the spatial derivative  that involves all lattice points, 
\begin{align}
\partial_x f(x,y)&\mapsto a^{-1}\sum_{n=1}^\infty(-1)^n n^{-1} [f(x-na,y)-f(x+na,y)]\nonumber\\
&=a^{-1}\sum_{n=1}^\infty(-1)^n n^{-1}(e^{-na\partial_x}-e^{na\partial_x})f(x,y)=a^{-1}(\ln e^{a\partial_x})f(x,y).\label{SLACderivative}
\end{align}
This discretization scheme goes by the name of {\sc slac} fermions \cite{Dre76,Cos02} in the particle physics literature. It has also been implemented in a condensed matter context \cite{Li18,Lan19,Lia22,Wan22}.

\begin{figure}[tb]
\centerline{\includegraphics[width=0.7\linewidth]{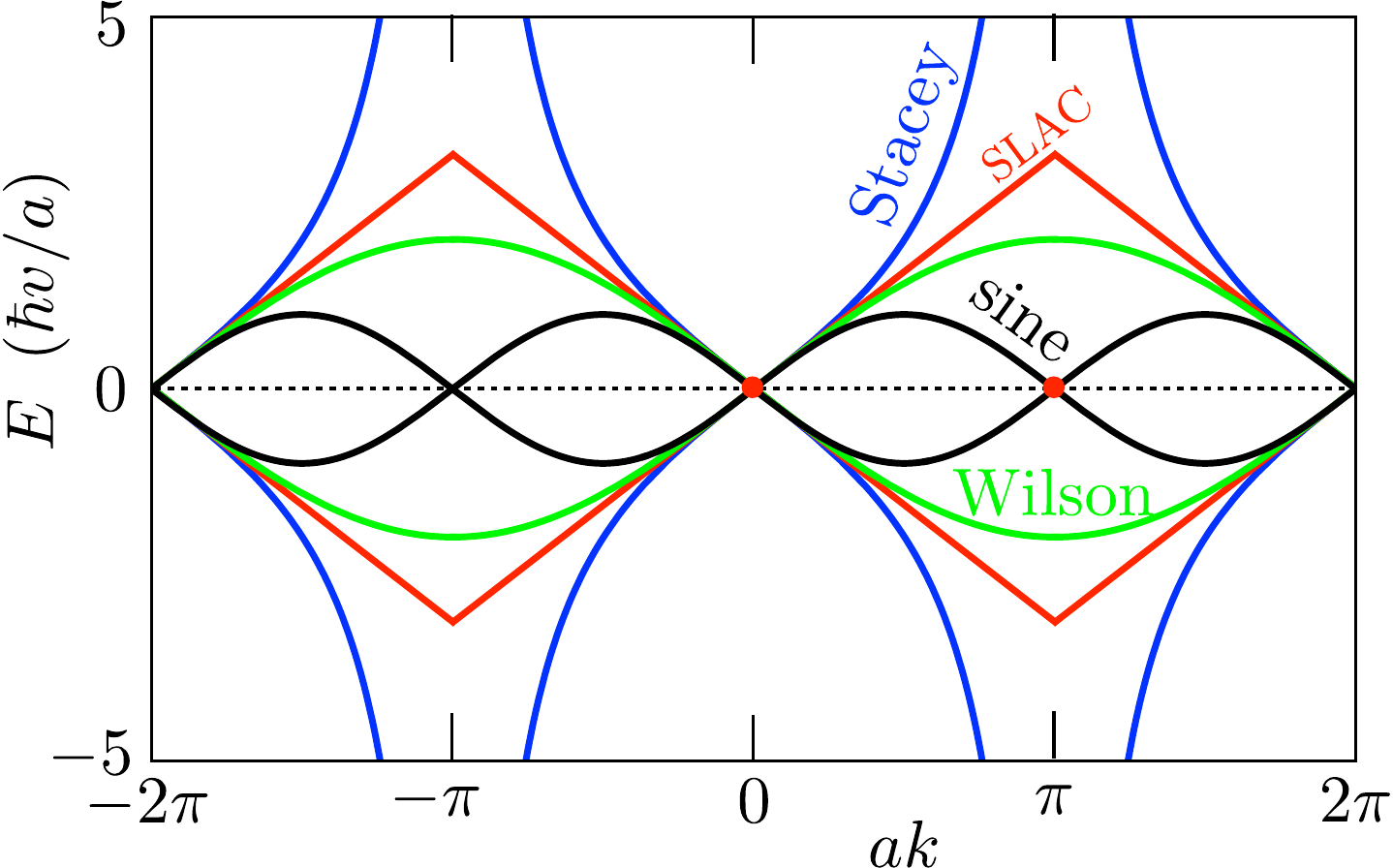}}
\caption{Dispersion relations of Dirac fermions on a 1D lattice, for four different discretization schemes. One with fermion doubling (black curve, $E_{\rm sine}$) and three without fermion doubling: $E_{\rm Wilson}$ (green curve, for $m_0=\hbar v/a$, when $E_{\rm Susskind}=E_{\rm Wilson}$), $E_{\rm SLAC}$ (red curve), and $E_{\rm Stacey}$ (blue curve). Inequivalent Dirac points are indicated by a red dot. The first Brillouin zone is the interval $|k|<\pi/a$, the plot is  extended to $|k|<2\pi/a$ to show the dispersion on both sides of the Brillouin zone boundary.
}
\label{fig_dispersion}
\end{figure}

In momentum representation, the Hamiltonian takes the form
\begin{equation}
H_{\rm SLAC}=-i(\hbar v/a)\bigl(\sigma_x \,\ln e^{iak_x}+\sigma_y \,\ln e^{iak_y}\bigr),
\end{equation}
where the branch cut of the logarithm is taken on the negative real axis. The corresponding dispersion 
\begin{equation}
E_{\rm SLAC}(\bm{k})^2=(\hbar v)^2\,(k_x^2+k_y^2)\;\;\text{for}\;\;|k_x|,|k_y|<\pi/a,\label{Ekslac}
\end{equation}
is a linear sawtooth, with a cusp at the edge of the Brillouin zone (see Fig.\ \ref{fig_dispersion}, red curve). Fermion doubling is avoided and both chiral and symplectic symmetries are preserved.

\subsection{Tangent dispersion}

The approach pioneered by Stacey \cite{Sta82,Ben83} seems a minor modification of the {\sc slac} approach --- but it has far reaching consequences. The nonlocal derivative \eqref{SLACderivative} is modified by removal of the $1/n$ factor, 
\begin{align}
\partial_x f(x,y)&\mapsto  2a^{-1}\sum_{n=1}^\infty (-1)^n  [f(x-na,y)-f(x+na,y)]\nonumber\\
&=2a^{-1}\sum_{n}(-1)^n (e^{-na\partial_x}-e^{na\partial_x})f(x,y)=-(2i/a)\tan(ia\partial_x/2)f(x,y).\label{Staceyderivative}
\end{align}
The corresponding Hamiltonian
\begin{equation}
H_{\rm Stacey}=(2\hbar v/a)\bigl[\sigma_x \,\tan(ak_x/2)+\sigma_y \tan(ak_y/2)\bigr],
\end{equation}
has a tangent dispersion, 
\begin{equation}
E_{\rm Stacey}(\bm{k})^2=(2\hbar v/a)^2\,\bigl[\tan^2 (ak_x/2)+\tan^2 (ak_y/2)\bigr].\label{Ekslacey}
\end{equation}
The cusp at the Brillouin zone boundary has been replaced by a pole (see Fig.\ \ref{fig_dispersion}, blue curve). 

As in the {\sc slac} approach, the Stacey approach avoids fermion doubling while preserving chiral and symplectic symmetries, at the expense of a nonlocal Hamiltonian. The key merit of the tangent dispersion is that the nonlocality can be removed by transforming the eigenproblem $H\Psi=E\Psi$ into a generalized eigenproblem ${\cal H}\Psi=E{\cal P}\Psi$, with local operators ${\cal H}$ and ${\cal P}$ on both sides of the equation. This transformation is possible because the tangent is the ratio of two operators, sine and cosine, that have a local representation on the lattice.\footnote{There is another way in which the Stacey discretization is local, which is that the inverse of the differential operation \eqref{Staceyderivative} is a local representation of the integral operator: the trapezoidal rule, see App.\ \ref{sec_trapezoidal}.}

Ref.\ \cite{Sta82} formulated the generalized eigenproblem by means of finite differences on a pair of staggered grids. This produces operators ${\cal H}$ and ${\cal P}$ that are local but not Hermitian, which is problematic in a numerical implementation. The alternative formulation of Ref.\ \cite{Pac21} resolves this issue, resulting in the generalized eigenproblem\footnote{For a real-space representation of Eq.\ \ref{genevproblem}, see App.\ \ref{sec_realspace}.}
\begin{equation}
\begin{split}
&{\cal H}\Psi=E{\cal P}\Psi,\;\;{\cal P}=\tfrac{1}{4}(1+\cos ak_x)(1+\cos ak_y),\\
&{\cal H}=\frac{\hbar v}{2a}\bigl[\sigma_x(1+\cos ak_y)\sin ak_x+\sigma_y(1+\cos ak_x)\sin ak_y\bigr].
\end{split}\label{genevproblem}
\end{equation}
Both operators ${\cal H}$ and ${\cal P}$ are Hermitian and ${\cal P}$ is also positive definite.\footnote{To avoid the complications from a noninvertible ${\cal P}$, one can choose a lattice with periodic boundary conditions over an odd number of sites; then all eigenvalues of ${\cal P}$ are strictly positive.} Both are sparse matrices, only nearby sites on the lattice are coupled. This combination of properties allows for an efficient calculation of the energy spectrum.

\section{Topologically protected Dirac point}

The Dirac cone of the continuum Hamiltonian \eqref{H0def} remains gapless in the presence of perturbations that do not break both chiral and symplectic symmetries. This topological protection is lost on the lattice for each of the discretized Hamiltonians discussed in the previous section --- except for one: Tangent fermions retain a topologically protected Dirac cone \cite{Don22b}.

The gap opening can be demonstrated in the simplest 1D case, in the presence of an electrostatic potential that varies rapidly on the scale of the lattice constant. This breaks chiral symmetry but it preserves symplectic symmetry, so in the continuum description the Dirac point should remain gapless (protected by Kramers degeneracy).

\begin{figure}[tb]
\centerline{\includegraphics[width=0.7\linewidth]{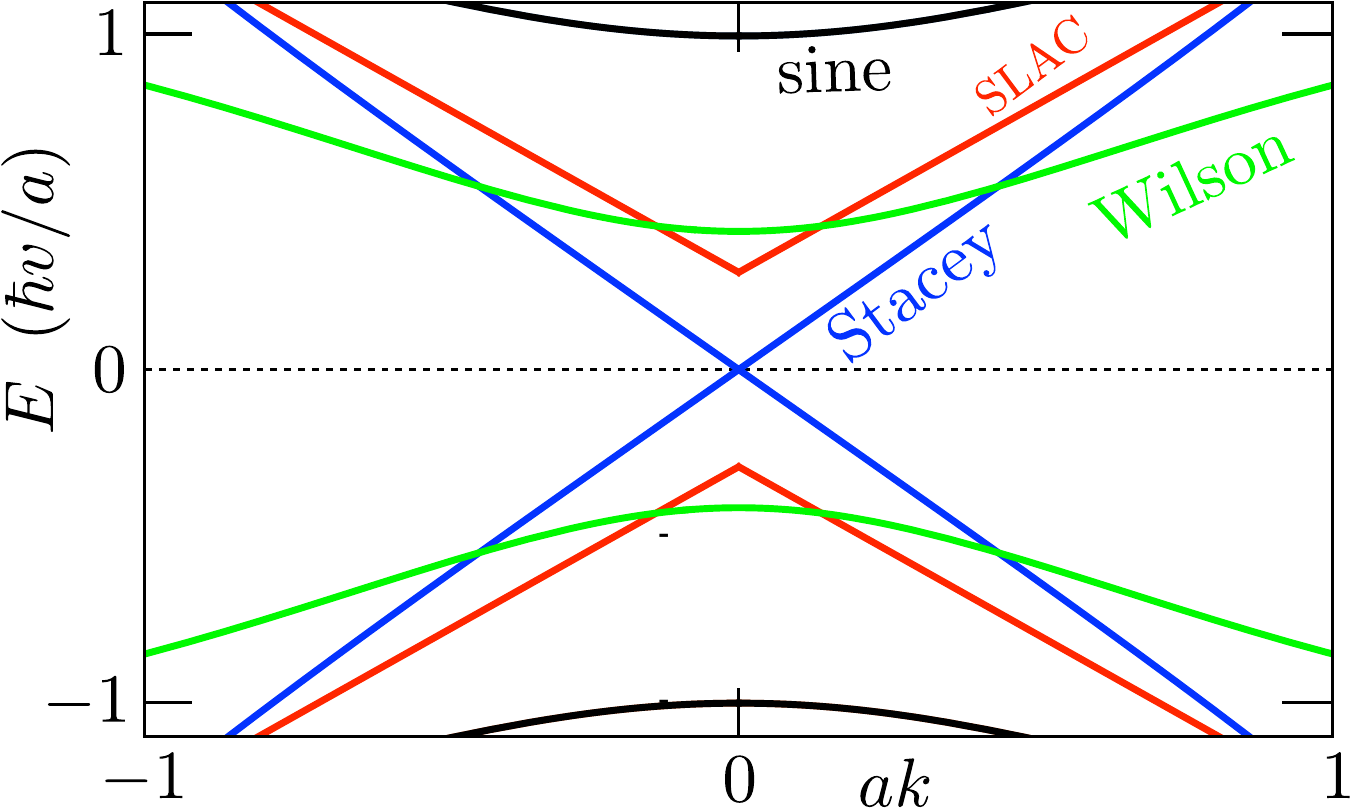}}
\caption{Same as Fig.\ \ref{fig_dispersion}, but now in the presence of a staggered potential $V(x)=V\cos(\pi x/a)$ with $V=\hbar v/a$. Only Stacey's tangent dispersion remains gapless.
}
\label{fig_gapopening}
\end{figure}

Following Ref.\ \cite{Don22b} we apply the staggered potential $V(x)=V\cos(\pi x/a)$, switching from $+ V$ to $-V$ between even and odd-numbered lattice sites. This potential couples the states at $k$ and $k+\pi/a$, as described by the Hamiltonian
\begin{equation}
H_{V}(k)=\begin{pmatrix}
H(k)&V/2\\
V/2&H(k+\pi/a)
\end{pmatrix}.
\end{equation}
The Brillouin zone is halved to $|k|<\pi/2a$, with the band structure\footnote{If $H(k)$ and $H(k+\pi/a)$ commute the band structure $E_V(k)$ in the presence of the staggered potential is given in terms of the unperturbed bandstructure $E(k)$ by $E_V(k)=\tfrac{1}{2}E(k)+\tfrac{1}{2}E(k+\pi/a)\pm\tfrac{1}{2}\sqrt{V^2+[E(k)-E(k+\pi/a)]^2}$. This applies to the sine, {\sc slac}, and Stacey discretizations. For the  Wilson and Susskind discretizations $E_V(k)$ has a more complicated expression.} shown in Fig.\ \ref{fig_gapopening}.

The size of the gap $\delta E$ that opens up at $k=0$ due to the staggered potential depends on the size of the gap $\Delta$ at $k=\pi/a$ of the unperturbed dispersion: the larger $\Delta$ the smaller $\delta E$. The sine dispersion has $\Delta=0$ and the resulting $\delta E=V$ is of first order in the perturbation strength. For the other dispersions the gap is of second order, $\delta E=V^2/\Delta$ for $V\ll \Delta$. To avoid the gap opening we thus need a pole $\Delta\rightarrow\infty$ in the dispersion at the Brillouin zone boundary, which is provided by the tangent dispersion.

\section{Application: Klein tunneling}

When a massless Dirac fermion approaches a potential barrier perpendicularly to the equipotentials it is not reflected but transmitted through the barrier with probability one. This effect, known as Klein tunneling, relies on two properties of the Dirac Hamiltonian: chiral symmetry and absence of fermion doubling \cite{All11,Bee08}. Reflection of the particle within the same Dirac cone would require a chirality flip and is thus forbidden.

\subsection{Tangent fermions on a space-time lattice}

Klein tunneling of tangent fermions was studied in Ref.\ \cite{Don22a}, based on a space-time lattice generalization \cite{Don22b} of the generalized eigenproblem \eqref{genevproblem}. The stationary equation ${\cal H}\Psi=E{\cal P}\Psi$ can be converted into a time-dependent equation (time step $\delta t$) upon substitution of $\Psi$ on the left-hand-side by $\tfrac{1}{2}[\Psi(t+\delta t)+\Psi(t)]$, and of $E\Psi$ on the right-hand-side by $(i\hbar/\delta t)[\Psi(t+\delta t)-\Psi(t)]$. 

The result is a finite difference equation of the Crank-Nicolson type,
\begin{equation}
\left({\cal P}+\frac{i\delta t}{2\hbar}{\cal H}\right)\Psi(t+\delta t)=\left({\cal P}-\frac{i\delta t}{2\hbar}{\cal H}\right)\Psi(t).\label{CNeq1}
\end{equation} 
Finite time step corrections are of third order in $\delta t$, but the evolution is exactly unitary to all orders in $\delta t$: $\Psi(t+\delta t)={\cal U}_0\Psi(t)$ with ${\cal U}_0=({\cal P}+\frac{i\delta t}{2\hbar}{\cal H})^{-1}({\cal P}-\frac{i\delta t}{2\hbar}{\cal H})$ unitary because ${\cal P}$ and ${\cal H}$ commute.

\begin{figure}[tb]
\centerline{\includegraphics[width=0.8\linewidth]{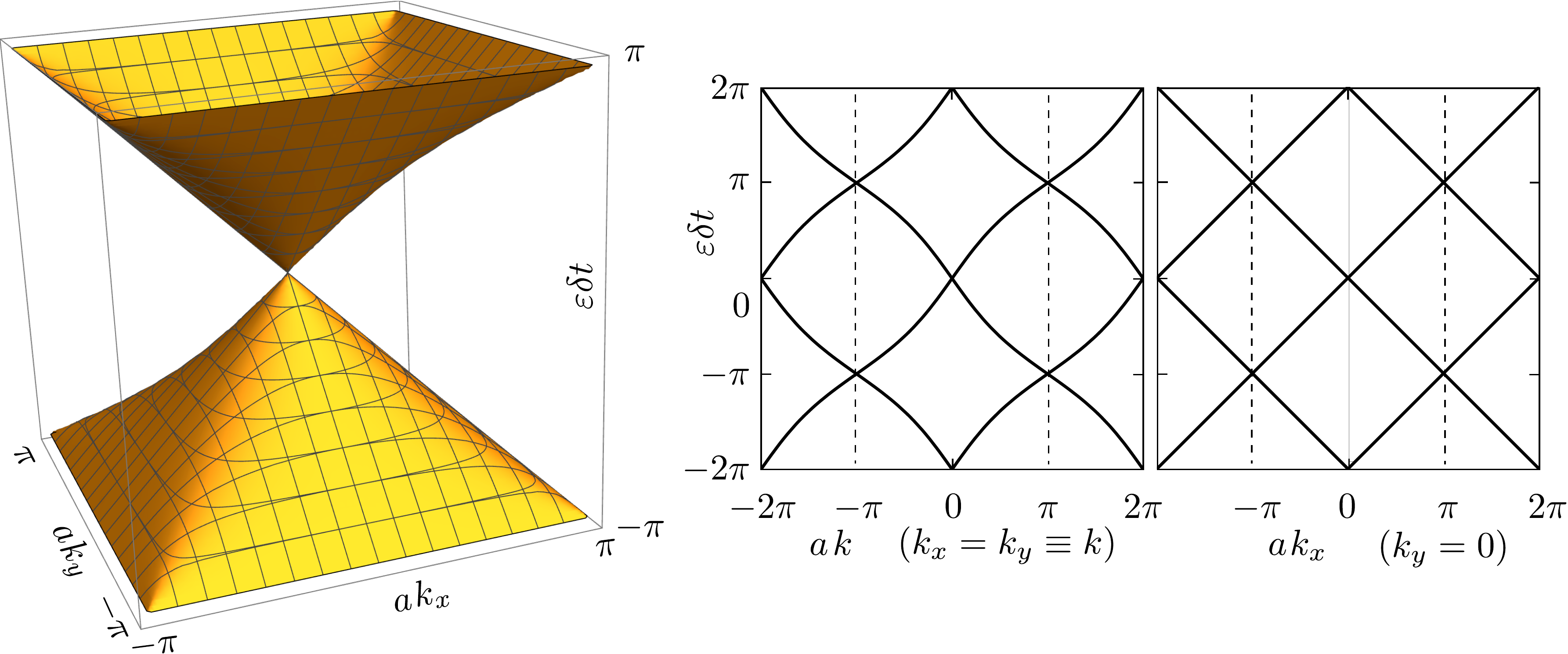}}
\caption{Band structure \eqref{dispersiontan} of tangent fermions on a space-time lattice (with $v\delta t/a=1$). The dispersion smoothly crosses the Brillouin zone boundaries (dotted lines). \textit{Figure from Ref.\ \cite{Don22b}.} \href{https://creativecommons.org/licenses/by/4.0/}{\tiny CC BY 4.0 license}}
\label{fig_dispersiontan}
\end{figure}

The eigenvalues $e^{i\varepsilon\delta t}$ of ${\cal U}_0$ are given by
\begin{equation}
\tan^2(\varepsilon\delta t/2)=(v\delta t/a)^2\left[ \tan^2(ak_x/2)+\tan^2(ak_y/2)\right],\label{dispersiontan}
\end{equation}
as plotted in Fig.\ \ref{fig_dispersiontan}. The dispersion is approximately linear near $\bm{k}=0$ and exactly linear along the lines $k_x=0$ and $k_y=0$ if we choose the discretization units such that $v=a/\delta t$. Alternatively, for $v=2^{-1/2}\,a/\delta t$ the dispersion is exactly linear along the diagonal lines $k_x=\pm k_y$.

Note that the pole in the tangent dispersion \eqref{Ekslacey} of the time-independent problem is regularized on the space-time lattice, the bands are joined smoothly at the Brillouin zone boundaries.

Eq.\ \eqref{CNeq1} describes the free evolution of the particle. For the Klein tunneling problem we wish to add an electrostatic potential $V(\bm{r})$. This can be done without breaking the unitarity of the evolution by splitting the operator \cite{Suz90},
\begin{equation}
e^{-i(H_0+V)\delta t/\hbar}=e^{-iV\delta t/2\hbar}e^{-iH_0\delta t/\hbar}e^{-iV\delta t/2\hbar} +{\cal O}(\delta t^3),
\end{equation}
resulting in the evolution equation \cite{Don22b}
\begin{equation}
\left({\cal P}+\frac{i\delta t}{2\hbar}{\cal H}\right)e^{iV\delta t/2\hbar}\Psi(t+\delta t)=\left({\cal P}-\frac{i\delta t}{2\hbar}{\cal H}\right)e^{-iV\delta t/2\hbar}\Psi(t).\label{CNeq2}
\end{equation}

Because both ${\cal P}$ and ${\cal H}$ are sparse matrices, coupling only nearby sites on the lattice, the finite difference equation \eqref{CNeq2} can be solved efficiently: The computational cost per time step scales as $N\ln N$ with the number of lattice sites. An alternative method of solution is to rewrite the evolution equation as
\begin{equation}
\Psi(t+\delta t)=e^{-iV\delta t/2\hbar}{\cal F}^{-1}{\cal U}_0{\cal F}e^{-iV\delta t/2\hbar}\Psi(t),\label{CNeq3}
\end{equation}
with ${\cal F}$ the Fourier transform operator. Since ${\cal U}_0$ is diagonal in momentum space and $V$ is diagonal in position space, the entire computational cost is then pushed into the fast Fourier transform algorithm, which has the same $N\ln N$ scaling.

\subsection{Wave packet propagation}

In Ref.\ \cite{Don22b} the evolution equation \eqref{CNeq3} was used to calculate the time dependence of a state $\Psi(x,y,t)$ incident along the $x$-axis on a rectangular barrier (height $V_0$). The initial state is a Gaussian wave packet,
\begin{equation}
\Psi(x,y,0)=(4\pi w^2)^{-1/2}e^{ik_0x}e^{-(x^2+y^2)/2w^2}{1\choose 1},\label{wavepacket}
\end{equation}
with parameters $k_0=0.5/a$, $w=30\,a$, at mean energy $\bar{E}=0.35\,\hbar/\delta t$. The velocity $v=2^{-1/2}\,a/\delta t$ was chosen such that the dispersion has the largest deviation from linearity along the $x$-axis, so this should provide the most stringent test of the space-time discretization.

\begin{figure}[tb]
\centerline{\includegraphics[width=0.8\linewidth]{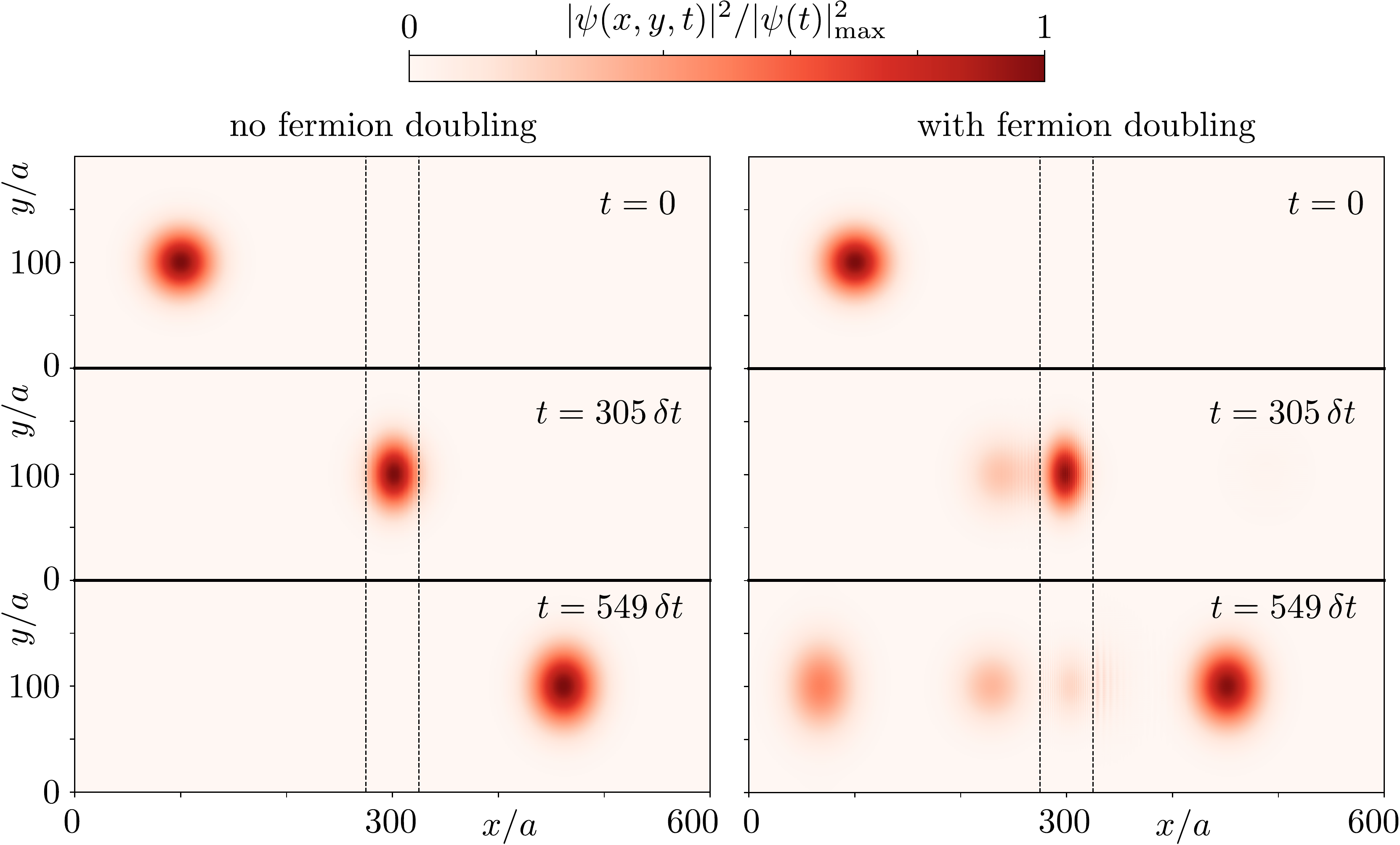}}
\caption{Three snapshots of the time-dependent simulation of Klein tunneling, in two alternative methods of discretization of the Dirac equation. A potential barrier (height $V_0=1.41\,\hbar/\delta t$  and width $50\,a$) is located between the dotted lines. A wave packet at mean energy $\bar{E}=0.35\,\hbar/\delta t$ is normally incident on the barrier. The color scale shows $|\Psi|^2$ normalized to unit peak height at each of the three times. Full transmission is obtained for the tangent discretization without fermion doubling (left panel), while fermion doubling causes reflections for a staggered space-time lattice discretization \cite{Ham14} (right panel). \textit{Figure from Ref.\ \cite{Don22a}.} {\tiny \copyright\ IOP Publishing. Reproduced with permission. All rights reserved.}
}
\label{fig_Klein1}
\end{figure}

As shown in the left panel of Fig.\ \ref{fig_Klein1}, although $V_0>\bar{E}$ the tangent fermion is fully transmitted through the potential barrier. This is contrasted in the right panel with the partial transmission of the wave packet for a discretization on a staggered space-time lattice \cite{Ham14}, which preserves chiral symmetry but has a second Dirac cone in the Brillouin zone.\footnote{The staggered discretization scheme of Ref.\ \cite{Ham14} is a rotation by $45^\circ$ of the Susskind lattice of Fig.\ \ref{fig_grids}, so the second Dirac cone is along the $k_x$-axis, and hence affects the Klein tunneling in the geometry of Fig.\ \ref{fig_Klein1}.}

\section{Application: Strong antilocalization}

One of the most striking signatures of an unpaired Dirac cone is the absence of quantum localization on the surface of a topological insulator \cite{Bar07,Nom07,Sbi14}. The conductance at the Dirac point \textit{increases} with increasing disorder, a counter-intuitive effect referred to as strong antilocalization (``strong'' to distinguish it from weak antilocalization, which is a small effect of order $e^2/h$). 

Disorder breaks chiral symmetry, but preserves the symplectic time-reversal symmetry --- which is essential for the effect. (In a magnetic field the surface would be localized by disorder, as in a quantum Hall insulator \cite{Eve08}.) Ref.\ \cite{Two08} studied the emergence of strong antilocalization in the electrical conductivity by application of Stacey's method of discretization of the Dirac equation \cite{Sta82,Ben83}. For that purpose the method must be applied to the transfer matrix rather than to the Hamiltonian. Let us summarize how that is done.

\subsection{Transfer matrix for tangent fermions}
\label{sec_transfermatrix}

We consider a two-terminal geometry along the $x$-axis, with a source contact at $x=0$ and a drain contact at $x=L$. To solve the scattering problem on a lattice we need to find out how the wave functions $\Psi(0,y)$ and $\Psi(L,y)$ are related in the presence of a disorder potential $V(x,y)$. 

We start from the generalized eigenproblem \eqref{genevproblem} for tangent fermions, including the potential in a way that preserves Hermiticity \cite{Pac21},
\begin{equation}
\begin{split}
&{\cal H}\Psi=\Phi^\dagger(E-V)\Phi\Psi,\;\;\Phi=\Phi_x\Phi_y,\;\;\Phi_\alpha=\tfrac{1}{2}(1+e^{iak_\alpha}),\\
&{\cal H}=\frac{\hbar v}{2a}\bigl[\sigma_x(1+\cos ak_y)\sin ak_x+\sigma_y(1+\cos ak_x)\sin ak_y\bigr].
\end{split}\label{genevproblem2}
\end{equation}
We have used the identity $\tfrac{1}{2}(1+e^{ik})(1+e^{-ik})=1+\cos k$ to factor the ${\cal P}$ operator in Eq.\ \eqref{genevproblem}. We then multiply both sides of the equation by $(\Phi_x^\dagger)^{-1}$, to obtain an equation that relates $\Psi_m(y)=\Psi(x=ma,y)$ to $\Psi_{m+1}=e^{iak_x}\Psi_m$:
\begin{align}
\Psi_{m+1}={}&\left[\Phi_y^\dagger(1+\tfrac{1}{2}i\sigma_x U_m)\Phi_y-\tfrac{1}{2}\sigma_z\sin ak_y\right]^{-1}\nonumber\\
&\cdot\left[\Phi_y^\dagger(1-\tfrac{1}{2}i\sigma_x U_m)\Phi_y+\tfrac{1}{2}\sigma_z\sin ak_y\right]\Psi_m,\label{Mmdef}
\end{align}
with $U_m=(a/\hbar v)[V(ma,y)-E]$.

Eq.\ \eqref{Mmdef} defines the one-step transfer matrix $M_m$, via $\Psi_{m+1}=M_m\Psi_m$. The full transfer matrix $M$ from source to drain is given by\footnote{The multiplication of transfer matrices is numerically unstable, because of the presence of both exponentially decreasing and increasing eigenvalues; the method to stabilize it is described in App.\ \ref{app_transmission}.} $\prod_m M_m$, ordered such that $M_{m+1}$ is to the left of $M_m$. The transfer matrix satisfies the current conservation condition
\begin{equation}
M^\dagger J_xM^{\vphantom{\dagger}}=J_x,\;\;J_x=v\sigma_x\Phi_y^\dagger\Phi_y.\label{currentconservation}
\end{equation}
The operator $J_x$ is the current operator in the $x$-direction. The transmission matrix $t$ can be obtained algebraically from the transfer matrix (see App.\ \ref{app_transmission}), and then the two-terminal conductance $G$ follows from the Landauer formula
\begin{equation}
G=\frac{e^2}{h}\,{\rm Tr}\,tt^\dagger,
\end{equation}
where $t$ is evaluated at the Fermi energy $E=E_{\rm F}$.

\subsection{Topological insulator \textit{versus} graphene}

\begin{figure}[tb]
\centerline{\includegraphics[width=0.5\linewidth]{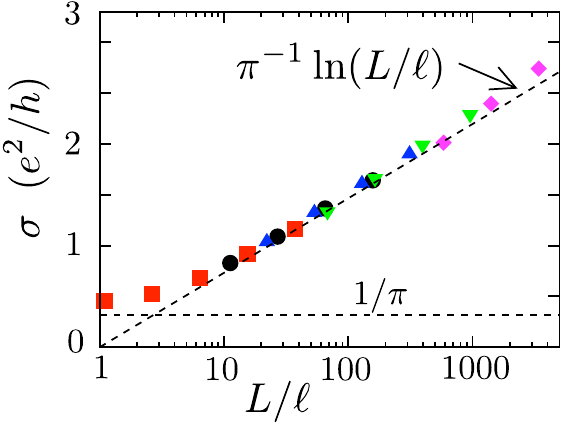}}
\caption{Dependence of the disorder averaged conductivity $\sigma=(L/W)\langle G\rangle$ on the ratio of sample length $L$ and mean free path $\ell$ of a topological insulator. Different colors of the data points distinguish different disorder strengths. The length $L$ and width $W$ of the sample are varied at constant aspect ratio $W/L=3$. The Fermi energy is at the Dirac point. The asymptotes expected in the limits of small and large $\ell$ are indicated by dashed lines. \textit{Figure from Ref.\ \cite{Two08}.}  {\tiny \copyright\ American Physical Society. Reproduced with permission. All rights reserved.} 
}
\label{fig_scale}
\end{figure}

In Fig.\ \ref{fig_scale} we show the scaling with sample size of the disorder-averaged surface conductivity $\sigma=(L/W)\langle G\rangle$ of a topological insulator \cite{Two08}. The conductivity at the Dirac point ($E_{\rm F}=0$) increases upon increasing the disorder strength, with a logarithmic scaling $\sigma\propto\ln(L/\ell)$ as a function of sample size $L$ and mean free path $\ell.$

\begin{figure}[tb]
\centerline{\includegraphics[width=0.6\linewidth]{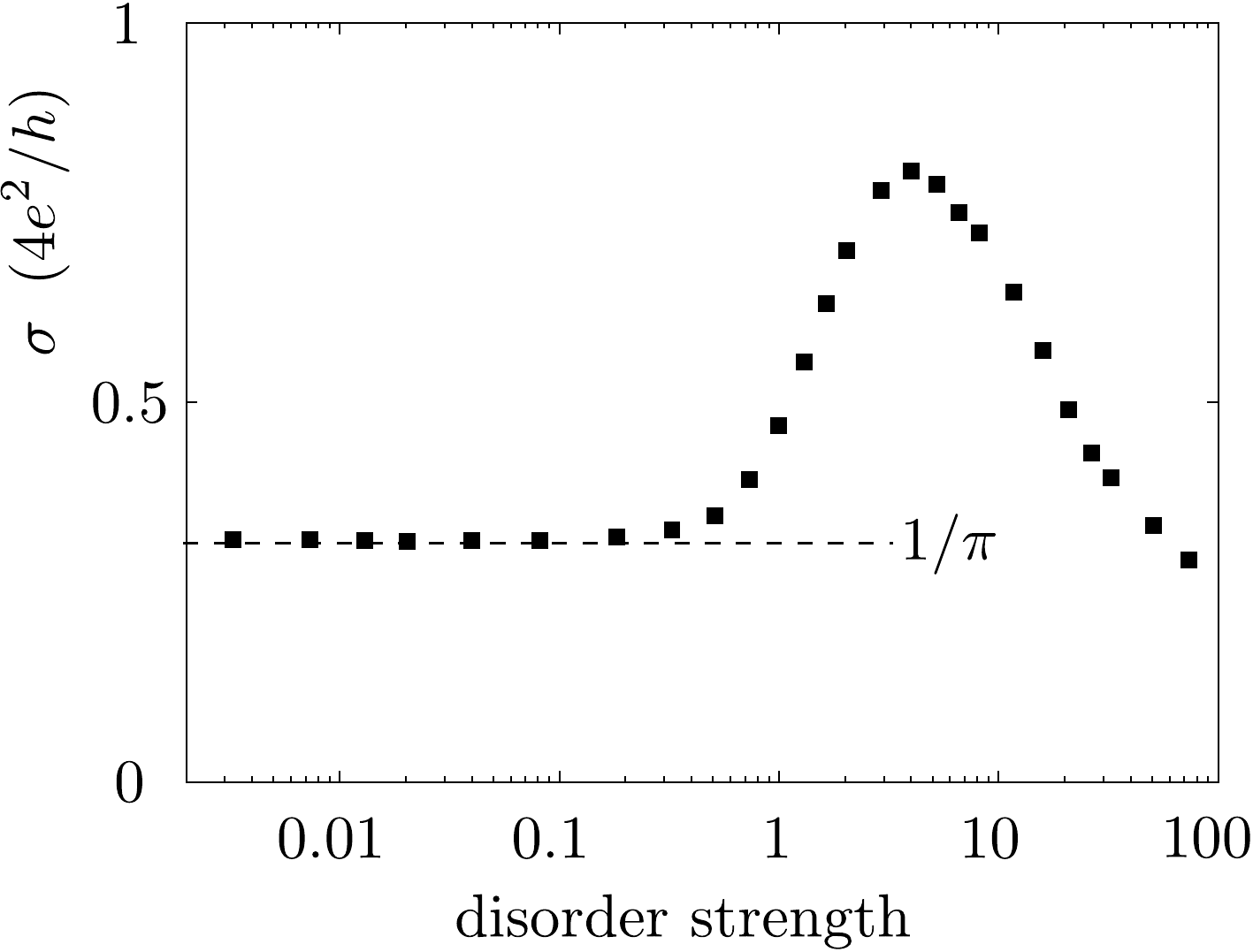}}
\caption{Disorder averaged conductivity of a graphene sheet ($W/L=3$, Fermi energy at the Dirac point), as a function of the dimensionless disorder strength. The conductance quantum is four times $e^2/h$, because of spin degeneracy and because of the presence of two Dirac points in the Brillouin zone. \textit{Figure from Ref.\ \cite{Ryc07}.  {\tiny \copyright\ European Physical Society. Reproduced with permission. All rights reserved}}
}
\label{fig_graphene}
\end{figure}

An unpaired Dirac cone is needed to avoid localization at large disorder. To see that, we show in Fig.\ \ref{fig_graphene} the conductivity as a function of disorder strength in a graphene sheet \cite{Ryc07}. Graphene has a 2D honeycomb lattice with a pair of Dirac cones in the Brillouin zone. A smooth and weak disorder potential does not couple the cones, so the conductivity initially increases with increasing disorder, as in a topological insulator. But when the disorder strength is further increased scattering between the Dirac cones of graphene becomes appreciable, and localization sets in.

\section{Application: Anomalous quantum Hall effect}

\begin{figure}[tb]
\centerline{\includegraphics[width=0.6\linewidth]{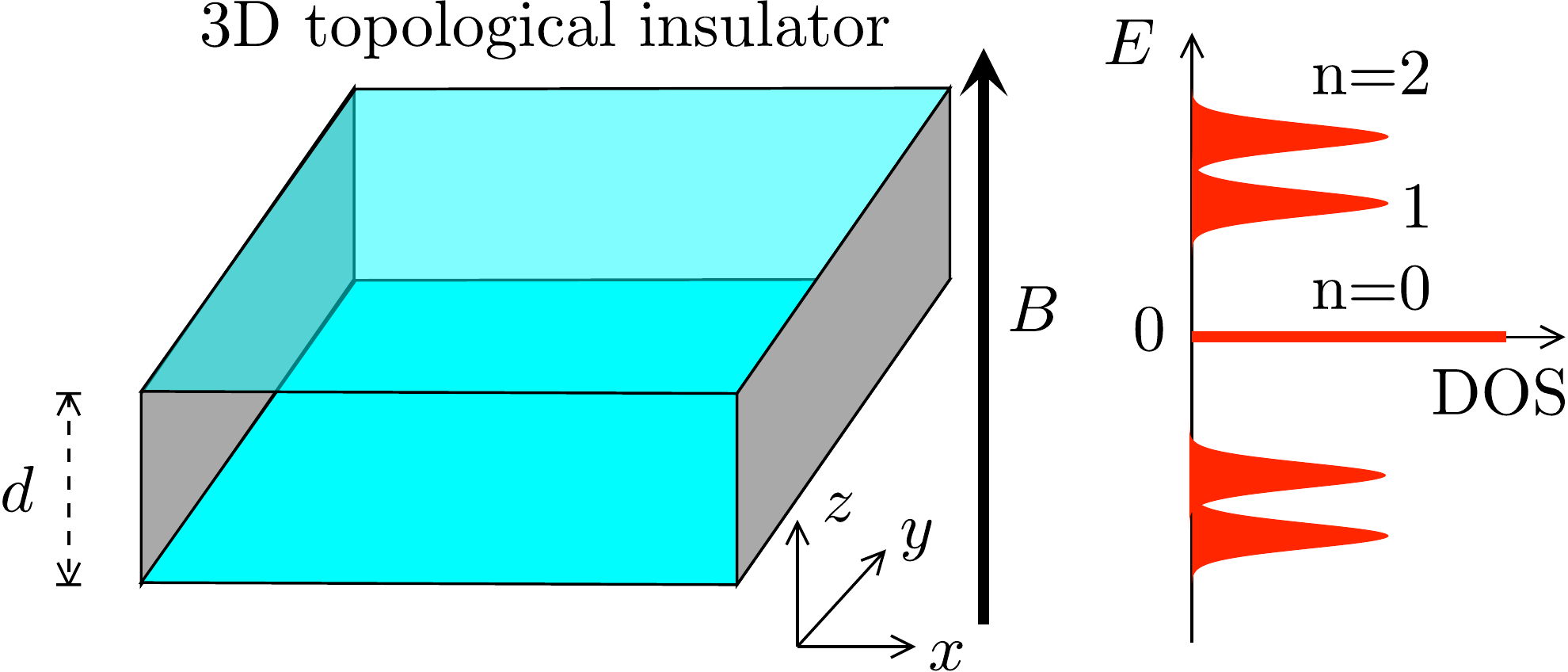}}
\caption{Slab of a topological insulator in a perpendicular magnetic field $B$. Landau levels form on the top and bottom surface at energy $|E|\propto\sqrt n$, $n=0,1,2,\ldots$, symmetrically arranged around $E=0$. The density of states (DOS) of the zeroth Landau level is not broadened by a spatially fluctuating $B$, provided that the slab thickness $d$ is sufficiently large that the two surfaces are decoupled. \textit{Figure from Ref.\ \cite{Don23}.} \href{https://creativecommons.org/licenses/by/4.0/}{\tiny CC BY 4.0 license}
}
\label{fig_anomalous}
\end{figure}

The spectrum of 2D massless Dirac fermions in a magnetic field is anomalous. In addition to magnetic field dependent Landau levels, there is one level pinned to zero energy irrespective of the magnetic field strength \cite{Kat12}. This ``zeroth Landau level'' is topologically protected by chiral symmetry: If the perpendicular field strength has spatial fluctuations, for example, because of ripples on the surface, all Landau levels are broadened \textit{except} the zeroth Landau level \cite{Gie07}. Fig.\ \ref{fig_anomalous} illustrates the anomaly.

To model this with tangent fermions on a 2D lattice we need to introduce the vector potential in a gauge invariant way --- without breaking the locality of the generalized eigenproblem. Let us summarize how that can be done, following Ref.\ \cite{Don23}.

\subsection{Gauge invariant tangent fermions}

We rewrite Eq.\ \eqref{genevproblem2} in terms of the translation operator $T_\alpha= e^{a\partial_\alpha}=e^{ia\hat{k}_\alpha}$,
\begin{subequations}
\label{genevproblem3}
\begin{align}
&{\cal H}\Psi=\Phi(E-V)\Phi^\dagger\Psi,\\
&{\cal H}=\frac{\hbar v}{8ia}\sigma_x(1+{T}_y)({T}_x-{T}_x^\dagger)(1+{T}_y^\dagger)+\frac{\hbar v}{8ia}\sigma_y(1+{T}_x)({T}_y-{T}_y^\dagger)(1+{T}_x^\dagger),\\
&\Phi=\tfrac{1}{8}(1+{T}_x)(1+{T}_y)+\tfrac{1}{8}(1+{T}_y)(1+{T}_x).
\end{align}
\end{subequations}

The Peierls substitution incorporates the vector potential $\bm{A}$ in a gauge invariant way by the replacement
\begin{equation}
T_\alpha\mapsto {\cal T}_\alpha=\sum_{\bm{n}}e^{i\phi_\alpha(\bm{n})}|\bm{n}\rangle\langle\bm{n}+\bm{e}_\alpha|,\;\;
\phi_\alpha(\bm{n})=e\int_{\bm{n}+\bm{e}_\alpha}^{\bm{n}}A_\alpha(\bm{r})\, dx_\alpha.
\label{Peierls}
\end{equation}
The sum over $\bm{n}$ is a sum over lattice sites on the 2D square lattice, and $\bm{e}_\alpha\in\{\bm{e}_x,\bm{e}_y\}$ is a unit vector in the $\alpha$-direction. Note that the $A$-dependent translation operators no longer commute,
\begin{equation}
{\cal T}_y{\cal T}_x=e^{2\pi i\varphi/\varphi_0}{\cal T}_x{\cal T}_y,
\end{equation}
where $\varphi$ is the flux through a unit cell in units of the flux quantum $\varphi_0=h/e$. In Eq.\ \eqref{genevproblem3} the translation operators are ordered such that ${\cal H}$ remains Hermitian.

This is the gauge invariant discretization of the time-independent Dirac equation. For the time-dependent case we apply the Peierls substitution \eqref{Peierls} to the finite difference equation
\begin{equation}
\left[\Phi\Phi^\dagger+\frac{i\delta t}{2\hbar}({\cal H}+\Phi V\Phi^\dagger)\right]\Psi(t+\delta t)=\left[\Phi\Phi^\dagger-\frac{i\delta t}{2\hbar}({\cal H}+\Phi V\Phi^\dagger)\right]\Psi(t).\label{CNeq4}
\end{equation} 
For $V=0$ this is the Crank-Nicolson equation \eqref{CNeq1}. Because $\Phi$ and ${\cal H}$ no longer commute after the Peierls substitution, the unitarity condition is modified. If we reorganize Eq.\ \eqref{CNeq4} as $\Psi(t+\delta t)={\cal U}\Psi(t)$, then the operator ${\cal U}$ satisfies the generalized unitarity condition
\begin{equation}
{\cal U}^\dagger \Phi\Phi^\dagger {\cal U}=\Phi\Phi^\dagger.
\end{equation}
It follows that the conserved density is \footnote{For the finite difference equation \eqref{CNeq2} the unitarity condition is ${\cal U}^\dagger{\cal U}=1$, with conserved density $\langle\Psi^\dagger|\Psi\rangle$, which is simpler, however, that unitarity breaks down if we include the vector potential $\bm{A}$. While the density \eqref{rhotdef} is conserved globally for any $\bm{A}$, a local formulation of the conservation law is only available to first order in $\bm{A}$ \cite{Pac21}.} 
\begin{equation}
\rho(t)=\langle\Psi^\dagger(t)|\Phi\Phi^\dagger|\Psi(t)\rangle=\rho(t+\delta t).\label{rhotdef}
\end{equation}

\subsection{Topologically protected zeroth Landau level}

In a continuum description the zeroth Landau level has a definite winding number, a chirality $C =\langle 0|\sigma_z|0\rangle=\pm 1$ depending on the sign of the magnetic field. If chiral symmetry is maintained the index $C$ is a topological invariant \cite{Kat12,Aha79,Alv83}, preventing a broadening of the level. However, on a lattice a nonzero $C$ is incompatible with gauge invariance \cite{Sta83}. Indeed, the Landau level spectrum for tangent fermions has a zeroth Landau level in both the $C=+1$ and $C=-1$ manifold \cite{Don23}.

One way to understand this obstruction, is to consider the process by which a uniform magnetic field is concentrated into an array of $h/e$ flux tubes, each of which is fully contained within a unit cell. The winding number cannot change by such a smooth deformation, but the resulting magnetic field distribution may be gauged away on the lattice, hence the net value of $C$ must be equal to zero. 

Fortunately, there is a work-around \cite{Don23}: One may spatially separate the opposite chirality manifolds by adjoining a $+B$ and $-B$ region next to each other in the 2D plane. Each of the two regions then has a chirality polarized zeroth Landau level. Nature employs a similar work-around in the 3D topological insulator geometry of Fig.\ \ref{fig_anomalous}, but there the chiralities are separated on opposite surfaces in the third dimension. The computational advantage of the tangent fermion discretization is that the spatial separation can be realized in two dimensions.

\begin{figure}[tb]
\centerline{\includegraphics[width=0.6\linewidth]{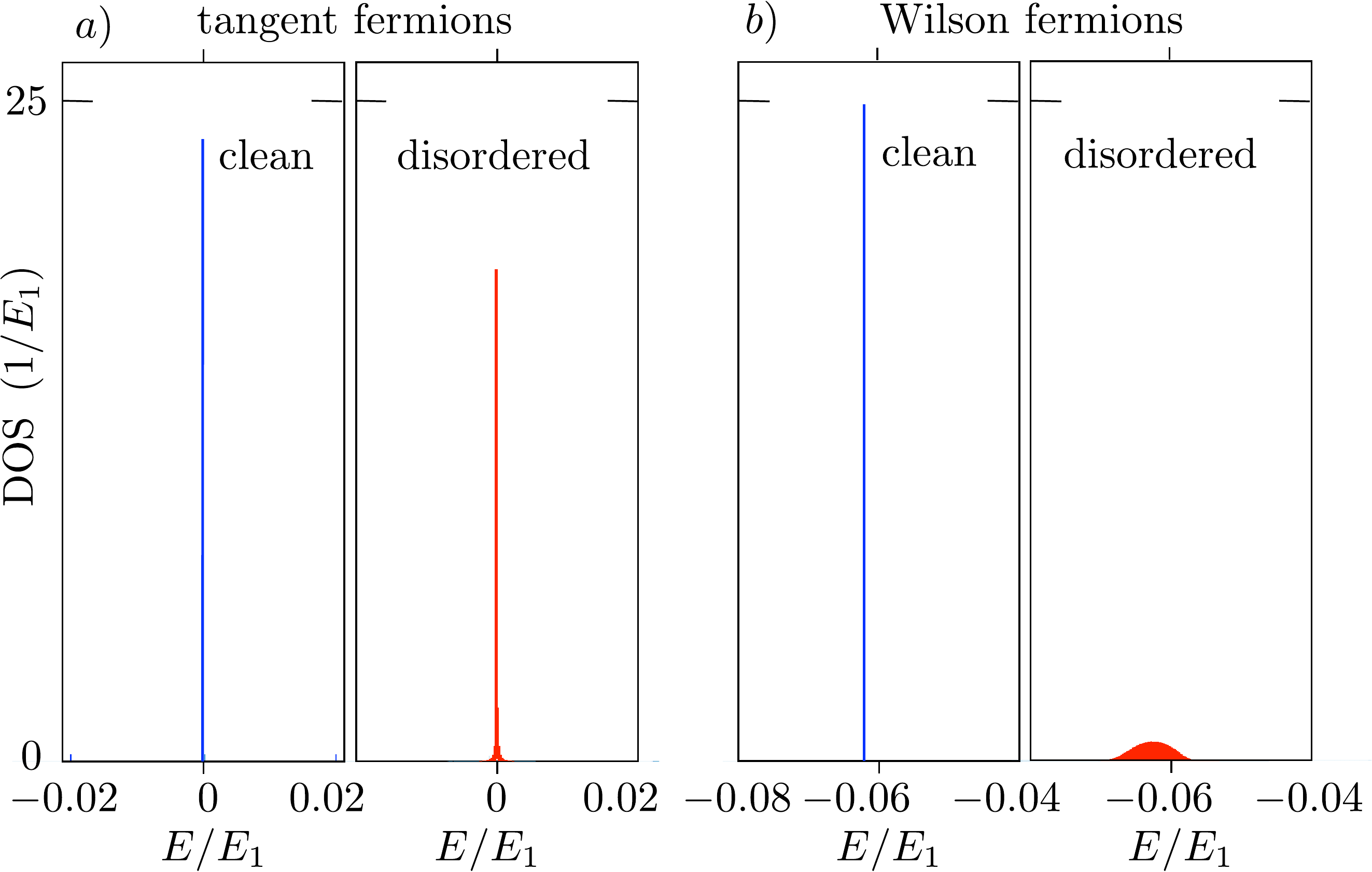}}
\caption{Density of states (DOS) per unit cell for the tangent dispersion (panel a) and for Wilson's sine+cosine dispersion (panel b), with and without disorder in the magnetic field. Energies are rescaled by the energy $E_1=v\sqrt{2\hbar eB}$ of the first Landau level. The zeroth Landau level of the Wilson Hamiltonian  \eqref{HWilson} is displaced from $E=0$ by an energy $\delta E=\tfrac{1}{2}eBa^2 m_0/\hbar$. The plot is for $m_0=\hbar v/a$. \textit{Figure from Ref.\ \cite{Don23}.} \href{https://creativecommons.org/licenses/by/4.0/}{\tiny CC BY 4.0 license}
}
\label{fig_DOS}
\end{figure}

The topological protection of the chirality-polarized zeroth Landau level is demonstrated in Fig.\ \ref{fig_DOS}. The left panel shows that the density of states peak at $E=0$ persists with only a slight broadening in the disordered system. It is essential that the tangent fermion discretization does not break chiral symmetry. To illustrate this, the right panel shows the corresponding result using Wilson's sine+cosine dispersion \eqref{EWilson}, which breaks chiral symmetry. Without disorder the only difference with the tangent dispersion is an energy shift of the zeroth Landau level \cite{Bru17,Bal18}, but with disorder the difference is quite dramatic. 

\section{Application: Majorana metal}

\subsection{Dirac versus Majorana fermions}

So far we discussed massless excitations, Dirac fermions, in a topological insulator. A topological superconductor also has massless excitations, but these are Majorana fermions rather than Dirac fermions \cite{Sen00,Rea00,Bee16}. The difference is the degree of freedom on which the Pauli matrices act in the $\bm{k}\cdot\bm{\sigma}$ Hamiltonian. For Dirac fermions the $\sigma_x$ operation flips the spin of the quasiparticle, for Majorana fermions it converts particle into antiparticle. The latter operation does not conserve charge, it is allowed because the missing charge of $2e$ is absorbed as a Cooper pair by the superconductor.

The Hamiltonian of a 2D topological superconductor is
\begin{equation}
H_{\rm Majorana}=\hbar v(k_x\sigma_x+k_y\sigma_y)+V(x,y)\sigma_z.
\end{equation}
In a topological insulator the electrostatic potential $V$ is multiplied by the unit matrix, since it does not couple to the spin of a Dirac fermion. But the potential acts with opposite sign on particles and antiparticles, hence the Pauli matrix $\sigma_z$ for Majorana fermions. The terms proportional to $\sigma_x$ and $\sigma_y$ represent a superconducting pair potential with \textit{p}-wave symmetry.

The chiral and symplectic symmetries \eqref{symmetries} are both broken, but the product of these two symmetries is preserved,
\begin{equation}
\sigma_x H_{\rm Majorana}^\ast\sigma_x=-H_{\rm Majorana},\label{sigmaxH}
\end{equation}
where the complex conjugation is carried out in the real-space basis (so momentum $\bm{k}$ changes sign). Eq.\ \eqref{sigmaxH} represents the particle-hole (or charge-conjugation) symmetry of a superconductor.

The particle-hole symmetry stabilizes localized states at zero energy: any small deviation away from $E=0$ would break the $\pm E$ symmetry. These socalled Majorana zero-modes are charge-neutral, equal-weight superpositions of electrons and holes. They cannot carry an electrical current, but they can contribute to thermal transport if their density becomes large enough. With increasing disorder the topological superconductor thus undergoes a transition from a thermal insulator to a thermal metal of Majorana fermions, a ``Majorana metal'' \cite{Sen00}.

\subsection{Phase diagram}

The properties of Majorana fermions on a lattice have been studied in Refs.\ \cite{Wim10,Med10}. One can use either the Wilson discretization approach (sine+cosine dispersion) or the Stacey approach (tangent dispersion), since the discretized Hamiltonian conserves particle-hole symmetry in both approaches. It is essential that fermion doubling is avoided: only an \textit{unpaired} Majorana zero-mode is stable at $E=0$, fermion doubling would allow the state to split away from zero energy without breaking the $\pm E$ symmetry.

\begin{figure}[tb]
\centerline{\includegraphics[width=0.7\linewidth]{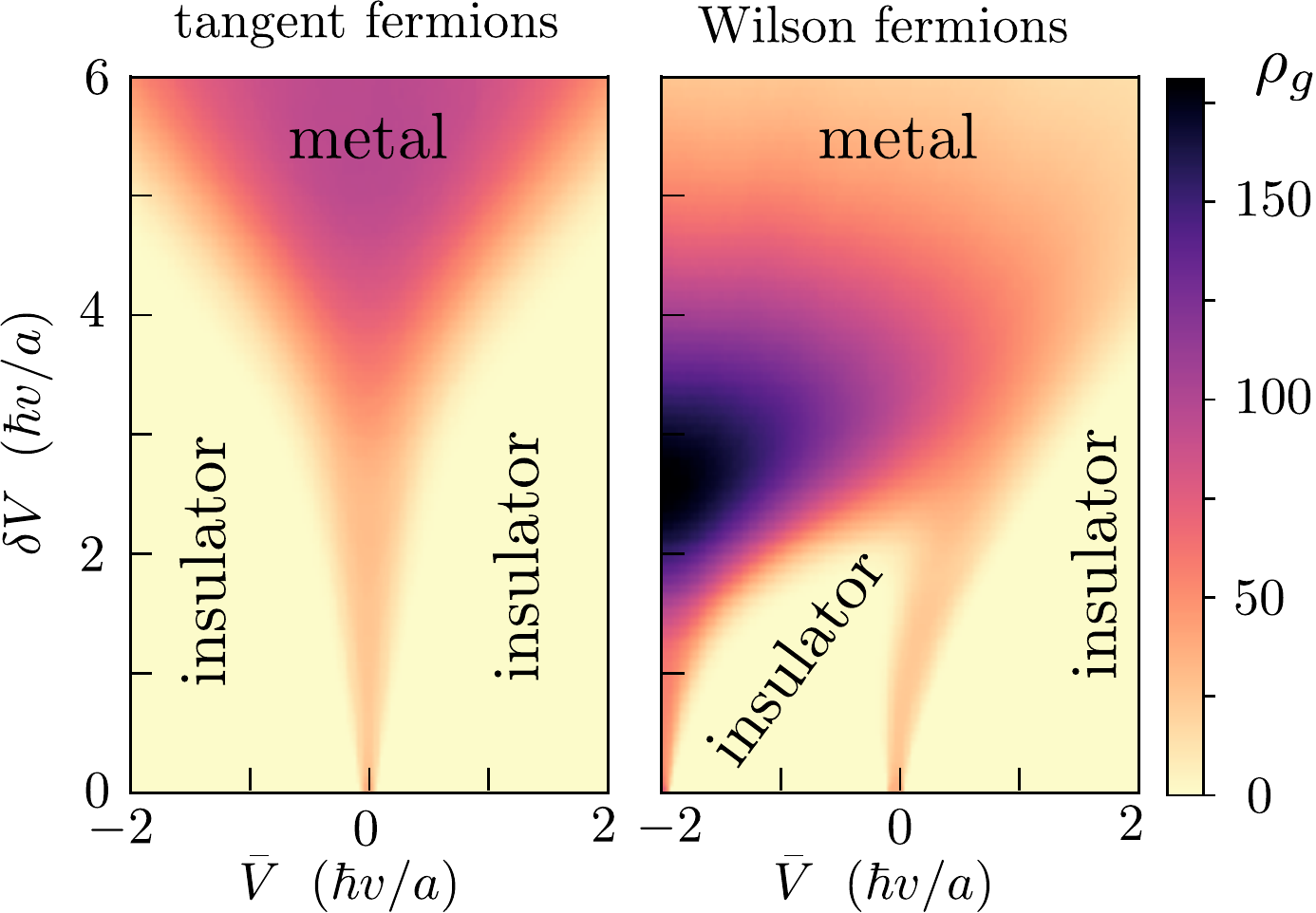}}
\caption{Phase diagram of Majorana fermions in a 2D topological superconductor. The potential landscape fluctuates randomly from site to site in the interval $(\bar{V}-\delta V,\bar{V}+\delta V)$. The color scale gives the geometrically averaged density of states $\rho_g$ \cite{Dob97,Jan98,Son07}, defined by $\ln (\rho_g/N)=N^{-1}\sum_{i=1}^N\log\rho_i$, with $\rho_i$ the local density of states at site $i\in\{1,2,\ldots N\}$ (computed in an energy window $\Delta E$ around $E=0$). States that extend over a subset $\delta N$ of the $N$ lattice sites have $\rho_g\simeq Ne^{-N/\delta N}$, so an exponentially small $\rho_g$ indicates a localized phase. The two plots are calculated for $N=20\times 20$, $\Delta E=0.23\,\hbar v/a$, averaged over 50 disorder realizations. The left panel is for tangent fermions, the right panel for Wilson fermions (with mass $m_0=\hbar v/a$).
}
\label{fig_Majorana}
\end{figure}

Fig.\ \ref{fig_Majorana} compares the phase diagrams for the two approaches. Both show the metallic phase at large disorder strength $\delta V$, but the insulating phase at weak disorder differs qualitatively. For tangent fermions the phase diagram is $\pm\bar{V}$ symmetric in the average potential. For the Wilson Hamiltonian \eqref{HWilson}, with mass term $m_0$, the symmetry axis is shifted: the phase diagram is $\pm(2m_0+\bar{V})$ symmetric. The phase boundary at $\bar{V}=0$ is thus replicated at $\bar{V}=-4m_0$ (outside of the range of the figure), and in between there is another phase boundary at $\bar{V}=-2m_0$. 

\section{Outlook}

Tangent fermions have not found much employ in lattice gauge theory \cite{Tong}. Richard Stacey, who introduced this discretization of the Dirac equation in that context \cite{Sta82}, concluded in a follow-up paper \cite{Sta85} that: ``This approach is not a success, and we will not consider it further.'' The singularity of the tangent dispersion at the edge of the Brillouin zone, which tangent fermions share with {\sc slac} fermions \cite{Dre76}, was considered a showstopper.

\begin{figure}[tb]
\centerline{\includegraphics[width=0.5\linewidth]{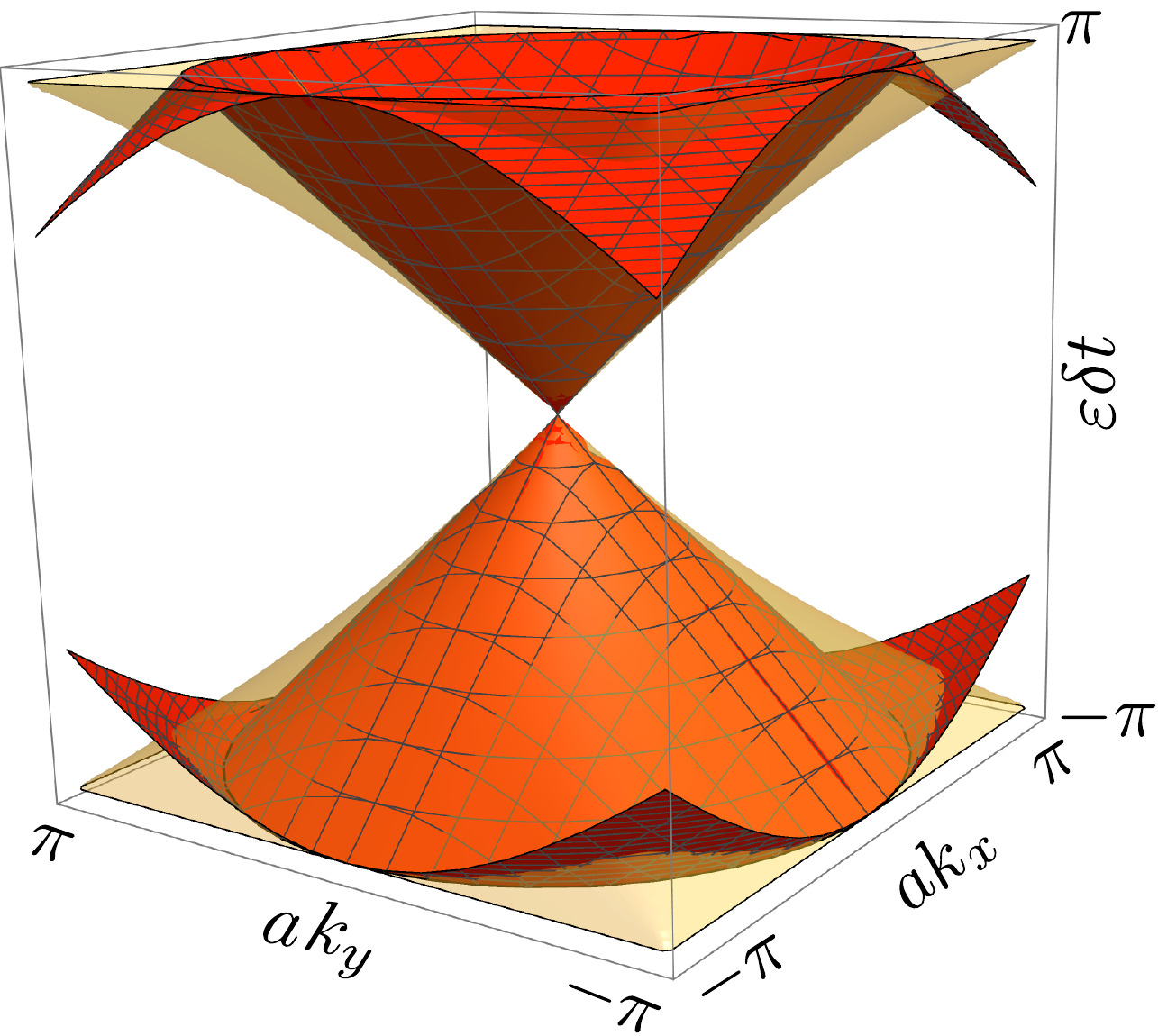}}
\caption{Quasi-energy bandstructure $\varepsilon(k_x,k_y)$ for {\sc slac} fermions (red) and for tangent fermions (yellow). The surfaces are computed, respectively, from the two equations $(\varepsilon\delta t+2\pi n)^2=(ak_x)^2+(ak_y)^2$, $n\in\mathbb{Z}$, and $\tan^2(\varepsilon\delta t/2)=\tan^2(ak_x/2)+\tan^2(ak_y/2)$. Only the first Brillouin zone is shown, the full bandstructure is periodic in momentum $k_\alpha$ with period $2\pi/a$ and periodic in quasi-energy $\varepsilon$ with period $2\pi/\delta t$. Near $\bm{k}=0$ both discretizations have the Dirac cone $\varepsilon^2=v^2(k_x^2+k_y^2)$ of the continuum limit, with velocity $v=a/\delta t$. The band structure of {\sc slac} fermions has a discontinuous derivative at Brillouin zone boundaries, the band structure of tangent fermions is smooth. \textit{Figure from Ref.\ \cite{Don22b}}. \href{https://creativecommons.org/licenses/by/4.0/}{\tiny CC BY 4.0 license}
}
\label{fig_BZdispersion}
\end{figure}

Stacey's approach was reconsidered in condensed matter physics, as a way to model the low-energy properties of graphene \cite{Two08}. The present review was motivated by a recent development in the study of topological insulators. Tangent fermions play a unique role in these materials, they represent \textit{the only} class of 2D lattice fermions with a topologically protected Dirac cone \cite{Don22b}. What distinguishes them from {\sc slac} fermions is that the dispersion can be regularized on a space-time lattice, producing a smooth energy-momentum relation across the entire Brillouin zone, of the form $\tan^2 (\varepsilon/2) =\tan^2 (k_x/2) +\tan^2 (k_y/2)$ in dimensionless units. The sawtooth dispersion of {\sc slac} fermions, in contrast, retains singularities at Brillouin zone boundaries when time and space are both discretized (see Fig.\ \ref{fig_BZdispersion}).

In the preceding sections we have reviewed several applications of tangent fermions in topological states of matter (insulators and superconductors). These all involve single-particle physics (either of Dirac fermions or of Majorana fermions). We look forward to applications in many-body physics. {\sc slac} fermions have been used to study various models of interacting electrons on a lattice \cite{Li18,Lan19,Lia22,Wan22}. Their sawtooth dispersion creates lattice artefacts which are not removed by reducing the lattice constant. It would be of interest to explore whether tangent fermions can provide an alternative route without those artefacts.

\section*{Acknowledgments}
C.B. received funding from the European Research Council (Advanced Grant 832256).\\
J.T. received funding from the National Science Centre, Poland, within the QuantERA II Programme that has received funding from the European Union's Horizon 2020 research and innovation programme under Grant Agreement Number 101017733, Project Registration Number 2021/03/Y/ST3/00191, acronym {\sc tobits}.

\appendix

\section{The inverse of the Stacey derivative is the trapezoidal integrator}
\label{sec_trapezoidal}

In Sec.\ \ref{sec_methods} we compared three different ways to discretize the differential operator $\partial_x$: the local derivative $\partial_x^{\text{local}}$ given by Eq.\ \eqref{sinediscrete}, the {\sc slac} derivative $\partial_x^{\text{SLAC}}$ given by Eq.\ \eqref{SLACderivative}, and the Stacey derivative $\partial_x^{\text{Stacey}}$ given by Eq.\ \eqref{Staceyderivative}. The corresponding inverses produce three different discretizations of the integral operator. Let us identify these.

Consider Riemann sum approximations of the integral $\int_{-\infty}^x f(x')\,dx'$ on a 1D lattice of equidistant points (spacing $a$). One distinguishes the right Riemann sum
\begin{equation}
S_+[f](x)=a\sum_{n=0}^\infty f(x-na)=\frac{a}{1-e^{-a\partial_x}},\label{S+def}
\end{equation}
the left Riemann sum
\begin{equation}
S_-[f](x)=a\sum_{n=0}^\infty f(x-a-na)=\frac{ae^{-a\partial_x}}{1-e^{-a\partial_x}},\label{S-def}
\end{equation}
and the middle Riemann sum
\begin{equation}
S_0[f](x)=2a\sum_{n=0}^\infty f(x-a-2na)=\frac{2ae^{-a\partial_x}}{1-e^{-2a\partial_x}}=\frac{a}{i\sin a\hat{k}_x}.\label{S0def}
\end{equation}
We have written these in terms of the translation operator $e^{\pm a\partial_x}f(x)=f(x\pm a)$, with $\hat{k}_x=-i\partial_x$ the momentum operator. The average of left and right Riemann sums is the trapezoidal integration rule,
\begin{equation}
T[f](x)=\tfrac{1}{2}S_+[f](x)+\tfrac{1}{2}S_-[f](x)=\tfrac{1}{2}a\frac{1+e^{-a\partial_x}}{1-e^{-a\partial_x}}=\frac{a}{2i}\,\text{cotan}\, (\tfrac{1}{2}a\hat{k}_x).\label{trapdef}
\end{equation}

We can now identify the inverse of $S_0$ with the sine dispersion of the local discretization \eqref{sinediscrete}, and the inverse of $T$ with the tangent dispersion of the Stacey discretization \eqref{Staceyderivative},
\begin{equation}
\begin{split}
&S_0^{-1}=\partial_x^{\text{local}},\\
&T^{-1}=\partial_x^{\text{Stacey}}.
\end{split}
\end{equation}
Remarkably enough, tangent fermions implement the trapezoidal integration rule.

What about {\sc slac} fermions? Inversion of the {\sc slac} derivative gives the integration operator kernel
\begin{equation}
K(x,x')=\frac{1}{2\pi}\int_{-\pi/a}^{\pi/a} e^{ik(x-x')}\frac{1}{ik}\,dk=\frac{1}{\pi}\text{Si}\,(\pi x/a-\pi x'/a),
\end{equation}
with $\text{Si}\,(x)$ the sine integral function. The discretized integral then becomes
\begin{equation}
\int_{-\infty}^x f(x')\,dx'\mapsto \frac{a}{\pi}\sum_{n=-\infty}^\infty \bigl[ \text{Si}\,(\pi x/a-\pi n)+\pi/2\bigr]f(na).\label{slacdef}
\end{equation}
Unlike the Riemann sums \eqref{S+def}--\eqref{trapdef}, the integrator \eqref{slacdef} is nonlocal: the definite integral $\int_{x_1}^{x_2} f(x)dx$ requires a summation over an infinite number of terms.

\section{Real-space formulation of the generalized eigenproblem}
\label{sec_realspace}

The generalized eigenproblem \eqref{genevproblem} of tangent fermions can be formulated in the position basis upon the substitution
\begin{equation}
e^{iak_\alpha}\mapsto\sum_{\bm{n}}|\bm{n}\rangle\langle\bm{n}+\bm{e}_\alpha|.
\end{equation}
The sum over $\bm{n}=n_x\bm{e}_x+n_y\bm{e}_y$, with $n_x,n_y\in\mathbb{Z}$, is a sum over lattice sites on the 2D square lattice (lattice constant $a$).

We thus have the equation ${\cal H}\Psi=E{\cal P}\Psi$, with on the left-hand-side the operator
\begin{subequations}
\label{Ddef}
\begin{align}
&{\cal H}=-\frac{i\hbar v}{a}{\bm D}\cdot\bm{\sigma},\;\;{\bm D}=(D_x,D_y),\\
&D_x=\tfrac{1}{8}\sum_{\bm n}\biggl(2|{\bm n}\rangle\langle{\bm n+\bm{e}_x}| 
+
|{\bm n}\rangle\langle{\bm n+\bm{e}_x+\bm{e}_y}|+
|{\bm n}\rangle\langle{\bm n+\bm{e}_x-\bm{e}_y}|
\biggr)-\text{H.c.},\\
&D_y=\tfrac{1}{8}\sum_{\bm n}\biggl(2|{\bm n}\rangle\langle{\bm n+\bm{e}_y}| 
+
|{\bm n}\rangle\langle{\bm n+\bm{e}_x+\bm{e}_y}|+
|{\bm n}\rangle\langle{\bm n+\bm{e}_y-\bm{e}_x}|
\biggr)-\text{H.c.},
\end{align}
\end{subequations}
and on the right-hand-side the operator ${\cal P}=\Phi^\dagger\Phi$ with
\begin{align}
\Phi={}&\tfrac{1}{4}\sum_{\bm n}\biggl(|{\bm n}\rangle\langle{\bm n}|+|{\bm n}\rangle\langle{\bm n+\bm{e}_x}|+|{\bm n}\rangle\langle{\bm n+\bm{e}_y}|+|{\bm n}\rangle\langle{\bm n+\bm{e}_x+\bm{e}_y}|\biggr).
\end{align}
The abbreviation H.c.\ stands for ``Hermitian conjugate''. Both operators ${\cal H}$ and ${\cal P}$ are local, only nearby lattice points are connected.

\begin{figure}[tb]
\centerline{\includegraphics[width=0.4\linewidth]{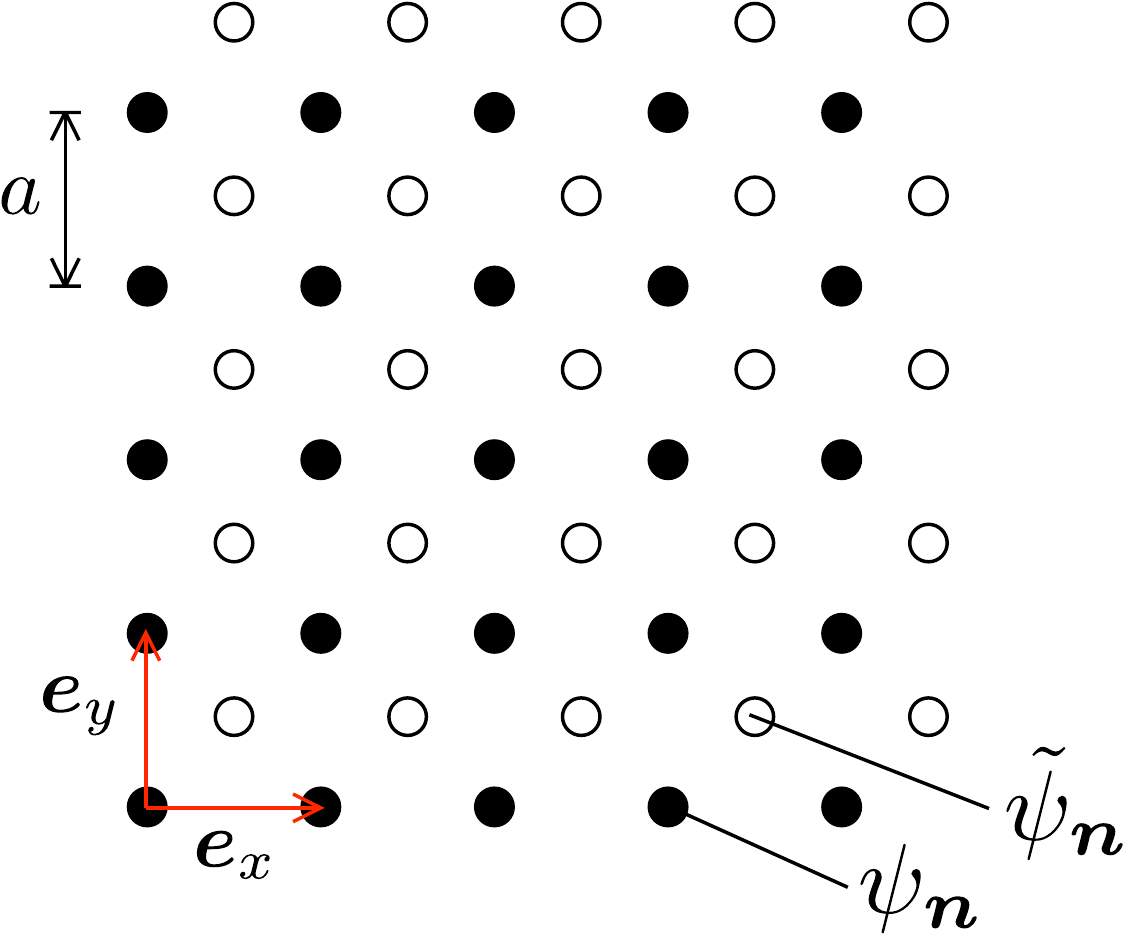}}
\caption{Staggered pair of grids to represent the two fields $\psi$ and $\tilde{\psi}=\Phi\psi$. \textit{Figure from Ref.\ \cite{Pac21}.} \href{https://creativecommons.org/licenses/by/4.0/}{\tiny CC BY 4.0 license}
}
\label{fig_staggered}
\end{figure}

By way of illustration, we work out the expectation value
\begin{equation}
\langle\psi|\Phi^\dagger \Phi|\psi\rangle=\sum_{\bm{n}}|\tilde{\psi}_{\bm{n}}|^2,\;\;\tilde{\psi}_{\bm{n}}=\tfrac{1}{4}(\psi_{\bm{n}}+\psi_{\bm{n}+\bm{e}_x}+\psi_{\bm{n}+\bm{e}_y}+\psi_{\bm{n}+\bm{e}_x+\bm{e}_y}).\label{psiPhidagPhipsi}
\end{equation}
One can interpret this in terms of the two staggered lattices shown in Fig.\ \ref{fig_staggered}. The field $\tilde{\psi}=\Phi\psi$ is defined on a white lattice point as the average of the amplitudes of the wave function $\psi$ on the four adjacent black lattice points.

\section{From transfer matrix to transmission matrix}
\label{app_transmission}

We explain the method \cite{Two08} to extract the transmission matrix $t$ from the transfer matrix $M$ of the tangent fermions. As described in Sec.\ \ref{sec_transfermatrix}, we consider a two-terminal geometry: A disordered region of width $W$ in the $y$-direction connects to ideal leads at $x=0$ and $x=L$. The dynamics in the leads is simplified by cutting the bonds in the transverse direction, so the incoming and outgoing modes propagate along one of the $N=W/a$ chains of lattice sites parallel to the $x$-axis.

We separate the spinor degrees of freedom $s=\pm$ of the transfer matrix into four $N\times N$ blocks,
\begin{equation}
{ M}=\begin{pmatrix}
{ M}^{++}&{ M}^{+-}\\
{ M}^{-+}&{ M}^{--}
\end{pmatrix}.\label{MsubblockB}
\end{equation}
The current conservation relation \eqref{currentconservation} can be written in the canonical form
\begin{equation}
\tilde{ M}^{\dagger}
\begin{pmatrix}
1&0\\
0&-1
\end{pmatrix}
\tilde{ M}=
\begin{pmatrix}
1&0\\
0&-1
\end{pmatrix},\label{Mcurrent2}
\end{equation}
in terms of a matrix $\tilde{ M}$ related to ${ M}$ by a similarity transformation,
\begin{equation}
\tilde{ M}={ R}{ M}{ R}^{-1},\;\;{ R}=\begin{pmatrix}
({\Phi_y^\dagger\Phi_y})^{1/2}&({\Phi_y^\dagger\Phi_y})^{1/2}\\
({\Phi_y^\dagger\Phi_y})^{1/2}&-({\Phi_y^\dagger\Phi_y})^{1/2}
\end{pmatrix}.\label{tildeM}
\end{equation}

It follows directly from Eq.\ \eqref{Mcurrent2} that the matrix $S$ constructed from $\tilde{ M}$ by
\begin{equation}
\tilde{ M}=\begin{pmatrix}
a&b\\
c&d
\end{pmatrix}
\Leftrightarrow
S=\begin{pmatrix}
-d^{-1}c&d^{-1}\\
a-bd^{-1}c&bd^{-1}
\end{pmatrix}\label{YtoU}
\end{equation}
is a unitary matrix. This is the scattering matrix of the disordered region. The $N\times N$ transmission matrix $t$ is the upper-right block of $S$, given by the inverse of the lower-right block of $\tilde{M}$,
\begin{equation}
t=\bigl(\tilde{ M}^{--}\bigr)^{-1}.
\end{equation}

The transformation \eqref{YtoU} also points the way to a method to avoid the numerical instability inherent in the multiplication of transfer matrices \cite{Two08}: First convert each $\tilde{M}_m$ into a unitary matrix $S_m$ by means of Eq.\ \eqref{YtoU}. Matrix multiplication of $\tilde{\cal M}_{m}$'s induces a nonlinear composition (star product \cite{Red59}) of $S_{m}$'s,
\begin{equation}
\tilde{\cal M}_{1}\tilde{\cal M}_{2}\mapsto S_{1}\star S_{2},\label{MMUU}
\end{equation}
defined by
\begin{align}
&\begin{pmatrix}
A_1&B_1\\
C_{1}&D_{1}
\end{pmatrix}
\star
\begin{pmatrix}
A_2&B_2\\
C_{2}&D_{2}
\end{pmatrix}\nonumber\\
&=\begin{pmatrix}
A_1+B_1(1-A_2D_1)^{-1}A_2C_1&B_1(1-A_2D_1)^{-1}B_2\\
C_2(1-D_1A_2)^{-1}C_1&D_2+C_2(1-D_1A_2)^{-1}D_1B_2
\end{pmatrix}.
\end{align}
Since the $S_m$'s are unitary, and the star product preserves unitarity, the divergence of eigenvalues that plagues the transfer matrix multiplication is avoided.

\end{document}